\documentclass[10pt]{article}

\usepackage{fullpage}
\usepackage{CJKutf8} 
\usepackage[utf8]{inputenc}
\usepackage{setspace}
\usepackage{parskip}
\usepackage{titlesec}
\usepackage[section]{placeins}
\usepackage{xcolor}
\usepackage{breakcites}
\usepackage{lineno}
\usepackage{hyphenat}
\usepackage{xcolor}

\usepackage{amsmath}
\usepackage{authblk}
\usepackage{amsfonts}
\usepackage{bm, amssymb}
\usepackage{mathtools}
\usepackage{graphicx}
\usepackage{subcaption}

\PassOptionsToPackage{hyphens}{url}
\usepackage[colorlinks = true,
            linkcolor = blue,
            urlcolor  = blue,
            citecolor = blue,
            anchorcolor = blue]{hyperref}
\usepackage{etoolbox}

\usepackage{natbib}

\renewenvironment{abstract}
  {{\bfseries\noindent{\abstractname}\par\nobreak}\footnotesize}
  {\bigskip}

\titlespacing{\section}{0pt}{*3}{*1}
\titlespacing{\subsection}{0pt}{*2}{*0.5}
\titlespacing{\subsubsection}{0pt}{*1.5}{0pt}

\usepackage{authblk}

\usepackage{graphicx}
\usepackage[space]{grffile}
\usepackage{latexsym}
\usepackage{textcomp}
\usepackage{longtable}
\usepackage{tabulary}
\usepackage{booktabs,array,multirow}
\usepackage{amsfonts,amsmath,amssymb}
\providecommand\citet{\cite}
\providecommand\citep{\cite}

\newif\iflatexml\latexmlfalse

\AtBeginDocument{\DeclareGraphicsExtensions{.pdf,.PDF,.eps,.EPS,.png,.PNG,.tif,.TIF,.jpg,.JPG,.jpeg,.JPEG}}

\usepackage[utf8]{inputenc}
\usepackage[english]{babel}
\usepackage{float}
\usepackage[margin=1.5in]{geometry}

\begin{document}
\begin{CJK}{UTF8}{gbsn}

\title{A Review of Bayesian Surrogate Modeling for High-Dimensional Computer Simulations}

\author[1]{Gavin Collins}
\author[1,4]{Gabriel Huerta}
\author[1,3]{J. Derek Tucker}
\author[2]{Kellin Rumsey}

\affil[1]{Statistical Sciences, Sandia National Laboratories}
\affil[2]{Statistical Sciences, Los Alamos National Laboratory}
\affil[3]{Department of Statistics, University of Illinois Urbana-Champaign}
\affil[4]{Department of Mathematics and Statistics, University of New Mexico}
\vspace{-1em}

\date{}

\begingroup
\let\center\flushleft
\let\endcenter\endflushleft
\maketitle
\endgroup

\begin{abstract}
Bayesian surrogate models play a central role in the analysis of computationally expensive physics-based simulators, particularly when uncertainty quantification (UQ) is a primary objective. This review provides a high-level overview of commonly used Bayesian surrogate modeling methods for high-dimensional computer simulations, with emphasis on both methodological foundations and practical applications. We review the advantages of Bayesian surrogates, the challenges posed by high-dimensional outputs, and two broad classes of methods: Gaussian process models, including scalable approximations, and Bayesian basis function regression approaches such as polynomial chaos expansions, Bayesian multivariate adaptive regression splines, Bayesian projection pursuit regression, and Bayesian additive regression trees. We also review strategies for handling high-dimensional simulator outputs, of which dimension-reduction approaches based on low-dimensional output representations are most common. Finally, we highlight key applications of Bayesian surrogate modeling in forward UQ, global sensitivity analysis, and Bayesian model calibration, and demonstrate a practical workflow utilizing these methods in simulated and real data analyses. We conclude by identifying open challenges and future research directions related to method selection, computational scalability, predictive accuracy, and UQ fidelity.
\end{abstract}

\section{Introduction} \label{sec:intro}

Physics-based computer simulations play a central role in modern science and engineering by providing a practical alternative when physical experiments are costly, dangerous, time-consuming, or simply infeasible. For example, climate and Earth system models are used to study past, present, and future climate behavior that cannot be manipulated experimentally on a global scale \citep{IPCC2021,Golaz2019}; computational fluid dynamics is used to analyze aerodynamic lift, drag, and stability characteristics before fabricating physical prototypes \citep{Anderson1999,Slotnick2014}; 
high-energy-density physics simulations are used to study matter under extreme temperatures and pressures that are difficult to reproduce experimentally \citep{Drake2018,Clark2016}; 
and nuclear engineering simulations are used to analyze reactor behavior and assess safety in situations where direct testing is restricted by cost or risk \citep{Williamson2012,Novascone2017}. In such settings, a simulator serves as a deterministic or stochastic map $y:\mathcal{X} \to \mathcal{Y},$ where $\bm x \in \mathcal{X}\subseteq \mathbb{R}^p$ denotes a vector of model inputs and $y(\bm x)$ denotes the corresponding simulator output.

Although modern simulators can accurately represent complex physical phenomena, and are often far more practical than conducting repeated physical experiments, their computational cost can become prohibitive when a large number of model evaluations is required. Depending on the complexity of the model and the degree of multiphysics coupling, a single simulation may take minutes, hours, or even days on high-performance computing platforms. Yet many scientific tasks require large ensembles of simulator evaluations over a range of inputs. This creates a significant bottleneck for downstream analyses such as uncertainty quantification (UQ), sensitivity analysis, optimization, calibration, and experimental design. As a result, there is strong motivation to replace expensive simulators with fast statistical approximations, or \emph{surrogate models}, that can emulate the simulator response with sufficient accuracy while reducing computational cost by orders of magnitude.

In this review, we focus exclusively on \emph{data-fit surrogates}, also called \emph{emulators}, \emph{metamodels}, or \emph{response surface models}. Unlike projection-based surrogates or lower-fidelity approaches, data-fit surrogates treat the simulator as a black-box function and learn an approximation from a set of input-output pairs, making them broadly applicable when the internal structure of the model is inaccessible \citep{smith2013}. Training inputs $\bm x_1,\dots,\bm x_n\in\mathcal{X}$ are selected using an experimental design strategy, with space-filling designs being a common choice, and the simulator is evaluated at these inputs to obtain responses $y(\bm x_1),\dots,y(\bm x_n)\in\mathcal{Y}$. A surrogate model is then selected and fit to the training data, and its predictive accuracy is evaluated using methods such as cross-validation. If performance is inadequate, one may revise the surrogate model or augment the training set with additional simulator runs. Once the surrogate is deemed adequate, it can be used in place of the expensive simulator for the intended inferential or decision-making task.

Many surrogate models are fit using maximum likelihood or other plug-in procedures that condition predictions on a single point estimate of the model parameters, and such approaches are often computationally attractive. In contrast, Bayesian methods place prior distributions on unknown quantities, update them using simulator data, and propagate posterior uncertainty into predictions and downstream analyses. This is especially important when the number of simulator runs is limited, or when the surrogate is embedded in a larger inferential workflow. In such settings, uncertainty in regression coefficients, covariance parameters, basis weights, or hyperparameters can substantially affect uncertainty in predicted responses, sensitivity analyses, posterior parameter distributions, or optimal decisions \citep{santner2003design}. Although some non-Bayesian methods also provide useful UQ, Bayesian methods offer a coherent framework for representing and propagating multiple sources of uncertainty. In this review, we limit our attention to Bayesian approaches.

Many simulators do not return a single scalar response, but instead produce high-dimensional outputs such as spatial fields, time series, spectra, or multiple quantities of interest observed across many locations, times, or conditions. A common strategy is to represent the output through a low-dimensional basis expansion, and fit independent surrogate models to the corresponding coefficient functions \citep{higdon2008computer}. In practice, the basis is often chosen using principal components, though other physically motivated representations are sometimes used. This reduces the problem of emulating high-dimensional output to a collection of scalar-output surrogate models and has become the dominant approach in practice. Other multivariate process models, such as the linear model of coregionalization \citep{JournelHuijbregts1978,FrickerOakleyUrban2013} and parallel partial Gaussian processes \citep{gu2016parallel}, provide a more direct way to model cross-output dependence, but are used less often because of their greater computational and modeling complexity.

The literature on surrogate modeling is vast, with several standard references providing broader background on surrogate modeling methodology, applications, and use cases \citep{santner2003design, FangLiSudjianto2006, ContiOHagan2010}. Our aim here is not to provide an exhaustive account, but rather to survey a substantial portion of the literature at a high level through the lens of Bayesian regression. In particular, we focus on two broad classes of methods. The first is Gaussian Process (GP) modeling, which places a stochastic process prior on the simulator response together with prior distributions on the hyperparameters of the covariance function; we also include commonly used GP approximations for large training datasets. The second is basis function regression, in which the response is represented through prescribed or learned basis expansions, with Bayesian priors placed on the associated coefficients, basis hyperparameters, and error terms. Although both can be interpreted within a common Bayesian regression framework, they involve different modeling choices and computational tradeoffs, and come with different inferential strengths and weaknesses. We do not attempt an exhaustive empirical comparison of these methods; instead, we highlight their principal strengths and weaknesses and demonstrate them on one simulated dataset and one real dataset. We refer readers to \cite{rumsey2025all} for a more comprehensive comparison.

The remainder of this article provides a high-level review of these ideas. Section~\ref{sec:surrogates} outlines the general surrogate modeling framework in the Bayesian context; then reviews common Bayesian GP models and Bayesian basis function regression approaches, highlighting their assumptions, strengths, limitations, and computational considerations; and finally describes extensions to simulations with high-dimensional output. Section~\ref{sec:use_cases} then discusses common use cases for Bayesian surrogate models, including forward uncertainty propagation, sensitivity analysis, and model calibration. To provide a practical anchor for the reviewed methods, Section~\ref{sec:demo} demonstrates their application to common surrogate modeling tasks on simulated multivariate-output data, and on a real dataset involving computational simulations coupled with experimental data from Sandia National Laboratories' Z machine. We conclude with a high-level summary, practical recommendations, and directions for future research in Section~\ref{sec:summary}.

\section{Surrogate Models}\label{sec:surrogates}
We begin by reviewing several widely used surrogate modeling approaches for approximating scalar-valued outputs from computer simulations. Our goal is to provide enough mathematical detail to convey the main ideas behind several common and effective Bayesian surrogate modeling approaches, while also giving a high-level overview of available software. For further details on implementation, model-specific choices, and practical nuances, we refer the reader to the broader literature. Although presented in the scalar-output setting, these methods provide the foundation for the high-dimensional output approaches discussed later in this section.

Let $y : \mathcal{X} \to \mathbb{R}$ denote a scalar-valued simulator response over an input space $\mathcal{X} \subseteq \mathbb{R}^p$. Suppose the simulator has been evaluated at design points $\bm{x}_1, \dots, \bm{x}_n \in \mathcal{X}$, yielding responses $y(\bm{x}_1), \dots, y(\bm{x}_n)$. The goal is to use these data to train a surrogate model $\hat{y} : \mathcal{X} \to \mathbb{R}$ that accurately approximates the simulator throughout the input space while providing a principled characterization of predictive uncertainty.

Throughout this review, we adopt the common modeling convention that the simulator response $y(\bm{x})$ at input $\bm{x}$ is represented as the sum of a surrogate approximation $\hat{y}(\bm{x})$ and a residual term $\epsilon(\bm{x})$:
\begin{equation}\label{eq:surr}
y(\bm{x}) = \hat{y}(\bm{x}) + \epsilon(\bm{x}).
\end{equation}

The residual error is then often modeled as
\begin{equation}
\epsilon(\bm{x}) \sim \mathcal{N}\!\left(0, v(\bm{x})\right),
\end{equation}
so that residual uncertainty is encoded in the variance function $v(\bm{x})$.

Bayesian surrogates provide a coherent framework for quantifying and propagating uncertainty in both $\hat{y}$ and $\epsilon$. This includes uncertainty arising from limited simulator evaluations, imperfect interpolation or extrapolation over the input space, and simulator noise when present. A principal advantage of the Bayesian approach is that uncertainty in model parameters and hyperparameters is propagated directly into predictive uncertainty, yielding a more complete characterization than classical methods based on maximum likelihood or related plug-in procedures \citep{Handcock1993, santner2003design}.

The surrogate models considered here fall into two broad classes: GP models and basis function regression models. These classes differ primarily in how they represent simulator response structure and quantify uncertainty. GPs model response covariance directly as a function of input proximity, and use this covariance structure for interpolation and UQ across $\mathcal{X}$. The variance function $v(\bm{x})$ induced by the GP posterior predictive distribution varies over the input space, with larger values at inputs $\bm{x}$ farther from the design points $\bm{x}_1, \dots, \bm{x}_n$. In contrast, basis function approaches approximate simulator output through potentially complex functional forms for the surrogate mean $\hat{y}(\bm{x})$, while the residual variance is often taken to be constant, $v(\bm{x}) = v > 0$ for all $\bm{x} \in \mathcal{X}$.

From a computational perspective, GPs offer a flexible characterization of uncertainty, but exact training requires inversion of an $n \times n$ covariance matrix, yielding computational cost that scales cubically with the number of training points. This can make exact GP inference impractical for large $n$, motivating the development of scalable approximation methods. Basis function methods are generally less flexible than full GPs in representing detailed covariance structure, but they are often more computationally efficient and can typically achieve comparable predictive accuracy.

While the present review centers on GP and basis function regression models, Bayesian neural networks \citep{neal1996bayesianlearning, Chang2021BayesianNN, Gawlikowski2021ASO} also warrant brief mention. Although neural-network-based surrogates can be highly expressive, we do not consider them further here because, in many practical settings, their UQ is not yet as reliable as that of GP and basis function approaches \citep{Bui2021BiasesIV, ries2022comparing, adams2023improving}. In addition, Bayesian inference for neural networks is often challenging due to the high-dimensional parameter spaces and complex dependence structures they entail \citep{pmlr-v139-izmailov21a}. For these reasons, we focus on methods whose UQ is better established in the computer experiments literature, while noting that Bayesian neural-network surrogates remain a promising and active area of research \citep{polson2026generative}.

In the following subsections, we first describe GP surrogates, including standard formulations and scalable approximations, and then present a general basis function regression framework together with several widely used variants. We later demonstrate these methods in a practical workflow for a simulated example in Section~\ref{sec:sim_example} and a real-data application in Section~\ref{sec:application}.

\subsection{Gaussian Process Regression}

GPs occupy a central role in the surrogate modeling literature for computer experiments, providing a flexible nonparametric representation of simulator response together with a natural probabilistic framework for quantifying predictive uncertainty \citep{oakley2002bayesian, santner2003design, rasmussen2006gpml}. For this reason, GP models have historically been the dominant surrogate modeling approach for expensive computer experiments \citep{sacks1989SS, currin1991bayesian, kennedy2001bayesian}, and they remain one of the most widely used methods in practice today \citep{gu2016parallel, plumlee2017JASA, gu2018scaled, gramacy2020surrogates}. We begin by introducing the basic GP model together with common formulations, then discuss Bayesian implementation of the exact GP, and finally review approximate GP approaches used to improve scalability.

A GP is a stochastic process for which, given any finite set of inputs $\bm{x}_1, \dots, \bm{x}_n \in \mathcal{X}$, the corresponding simulator outputs $y(\bm{x}_1), \dots, y(\bm{x}_n)$ have a joint multivariate normal distribution. It is fully characterized by a mean function $m : \mathcal{X} \to \mathbb{R}$, with $m(\bm{x}) = \mathbb{E}[y(\bm{x}) \mid \bm{x}]$, and a symmetric positive semidefinite covariance function, or kernel, $k : \mathcal{X} \times \mathcal{X} \to \mathbb{R}$, such that $k(\bm{x}, \bm{x}^\star) = \operatorname{Cov}(y(\bm{x}), y(\bm{x}^\star))$ for $\bm{x}, \bm{x}^\star \in \mathcal{X}$. The mean function is often taken to be constant, $m(\bm{x}) = m \in \mathbb{R}$ across $\mathcal{X}$, but we retain the more general notation for completeness. In practice, most of the modeling flexibility is typically introduced through the covariance function $k$, which is parameterized by a set of hyperparameters.

In surrogate modeling for computer experiments, a common starting point is a stationary covariance kernel, for which the covariance depends only on the separation between input locations. Anisotropic kernels with dimension-specific lengthscale parameters are typically used, allowing the response to vary at different rates across input dimensions. A standard formulation is
\begin{equation}
k(\bm x, \bm x^\star) = \sigma^2 \phi\left(\sqrt{\sum_{j=1}^p \frac{(x_j-x_j^\star)^2}{\ell_j^2}}\right),
\end{equation}
where $\sigma^2$ is a variance parameter, $\ell_j$ are dimension-specific lengthscales, and $\phi$ is a valid correlation function. Common choices for $\phi$ include members of the \emph{Matérn} family, which provides explicit control over smoothness. The exponential kernel is a low-smoothness Matérn case, while the squared exponential kernel can be viewed as the infinitely smooth limiting case.

When simulator output is stochastic, or when additional small-scale variation is needed for numerical reasons, a nugget term is commonly included:
\begin{equation}\label{eq:aniso_nugget}
k(\bm x, \bm x^\star) = \sigma^2 \phi\left(\sqrt{\sum_{j=1}^p \frac{(x_j-x_j^\star)^2}{\ell_j^2}}\right) + \tau^2 \bm{1}_{\bm x=\bm x^\star},
\end{equation}
where $\bm{1}_{\bm x=\bm x^\star}$ denotes the Kronecker delta, equal to 1 when $\bm x = \bm x^\star$ and 0 otherwise.

Although stationary anisotropic kernels of this form are often effective defaults, they are not the only option. Depending on the application, one may instead use isotropic kernels, nonstationary covariance structures, or heteroskedastic models that allow the noise variance to vary over the input space.

We now describe the form of the surrogate approximation $\hat{y}$ and residual variance $v(\bm x)$ after conditioning on the observed input-response pairs. For an observed set of computer model responses $\bm y = \left(y(\bm x_1), \dots, y(\bm x_n)\right)^\top$, let $\bm m = \left(m(\bm x_1), \dots, m(\bm x_n)\right)^\top$ and $K = \left[k(\bm{x}_i, \bm{x}_j)\right]_{i,j=1}^n$ denote the corresponding mean vector and covariance matrix, such that

\begin{equation}
    \bm y \sim \mathcal{N}(\bm m, K).
\end{equation}

To predict at a new input $\bm x^\star$, define $\bm k_\star = \left(k(\bm x^\star, \bm x_1), \dots, k(\bm x^\star, \bm x_n)\right)^\top$ and $k_{\star\star} = k(\bm x^\star, \bm x^\star)$. Conditional on $m$ and $k$, the GP model yields the predictive mean

\begin{equation}
    \mathrm{E}[y(\bm x^\star)\mid \bm y] = m(\bm x^\star) + \bm k_\star^\top K^{-1}(\bm y - \bm m)
\end{equation}
and predictive variance

\begin{equation}
    \mathrm{Var}(y(\bm x^\star)\mid \bm y) = k_{\star\star} - \bm k_\star^\top K^{-1} \bm k_\star.
\end{equation}

In the notation of Equation \eqref{eq:surr}, the predictive mean serves as the surrogate approximation $\hat{y}(\bm x^\star)$, while the predictive variance corresponds to the variance function $v(\bm x^\star)$, thereby providing both a flexible nonparametric approximation and principled UQ. The expressions above are the standard predictive equations conditional on the mean and covariance functions. In practice, however, these functions depend on unknown hyperparameters. The next subsection addresses how these quantities are estimated and how estimation uncertainty is propagated within a Bayesian framework.

\subsubsection{Bayesian Inference for Exact Gaussian Processes}

In the preceding discussion, the mean function $m$ and covariance function $k$ were written as if their hyperparameters were known. While it is common to assume a constant, zero mean $m(\bm x)=0$, the covariance function $k$ typically depends on an unknown parameter vector $\bm \theta$. For the kernel in Equation~\eqref{eq:aniso_nugget}, for example, we have

\begin{equation}
    \bm\theta = (\sigma^2, \ell_1, \dots, \ell_p, \tau^2).
\end{equation}

A fully Bayesian GP treats these quantities as unknown and assigns prior distributions to them, rather than relying on point estimation procedures such as maximum likelihood estimation. Given a prior distribution $p(\bm\theta)$, inference is based on the posterior distribution

\begin{equation}
    p(\bm\theta \mid \bm y) \propto p(\bm y \mid \bm\theta)\, p(\bm\theta),
\end{equation}
where $p(\bm y \mid \bm\theta)$ is the GP likelihood. The posterior predictive distribution at a new input $\bm x^\star$ is then obtained by averaging the usual GP predictive distribution over the posterior distribution of $\bm\theta$, so that uncertainty about the covariance and noise parameters is reflected in the resulting predictive uncertainty. These effects can be important for downstream tasks such as propagating uncertain inputs, sensitivity analysis, and calibration. Fully Bayesian inference can also lead to better-calibrated prediction intervals, especially when $n$ is small or the input space is high-dimensional \citep{santner2003design}, and supports direct probabilistic inference on hyperparameters, such as comparing lengthscale parameters across inputs to assess their relative importance.

Prior specification for $\bm\theta$ is an important part of Bayesian GP modeling. In practice, priors are often chosen independently across parameters as a starting point \citep{Paulo2005,santner2003design}. For variance and lengthscale parameters, common choices include log-normal, gamma, inverse-gamma, and half-$t$ priors \citep{Paulo2005,Gelman2006}. In some formulations, a conjugate inverse-gamma prior can be leveraged for the marginal variance parameter $\sigma^2$, allowing analytical marginalization. In many modern implementations, however, inference is carried out directly on transformed unconstrained parameters.

Markov chain Monte Carlo (MCMC) methods such as Metropolis-Hastings and Hamiltonian Monte Carlo can be used to obtain samples from the posterior distribution of $\bm \theta$, thereby enabling full propagation of hyperparameter uncertainty. Their main drawback is computational cost, since each posterior evaluation requires inversion of the covariance matrix $K$, which scales cubically with $n$ for exact GPs. This computational burden has motivated the development of a variety of approximation methods, several of which are described in the next subsection.

A wide range of software supports Bayesian GP modeling. General-purpose probabilistic programming frameworks such as \texttt{Stan} \citep{carpenter2017stan} and \texttt{PyMC} \citep{abrilpla2023pymc} allow custom models with user-specified priors and MCMC or variational inference. Dedicated GP libraries, including \texttt{GPyTorch} \citep{gardner2018gpytorch}, \texttt{GPflow} \citep{matthews2017gpflow}, and related Python tools, also support Bayesian inference to varying degrees. In \textsf{R}, packages such as \texttt{RobustGaSP} \citep{gu2018robustgasp}, \texttt{hetGP} \citep{binois2018hetgp}, \texttt{spareGParts} \citep{spareGParts}, \texttt{plgp} \citep{gramacy2011particle}, \texttt{tgp} \citep{gramacy2007tgp}, and \texttt{gplite} \citep{gplite} are commonly used for surrogate modeling and computer experiments. In practice, however, most GP packages emphasize optimization-based hyperparameter estimation, so full posterior inference often requires custom implementation or the use of a probabilistic programming framework.

\subsubsection{Approximate Gaussian Processes}

Although GP surrogates are a powerful tool for interpolation and UQ, they quickly become computationally burdensome as the number of training points $n$ increases. The main cost comes from factorizing or inverting the $n \times n$ covariance matrix $K$, which requires $\mathcal{O}(n^3)$ operations. In Bayesian settings, this burden is amplified because these covariance operations must be repeated many times across posterior samples of the kernel hyperparameters. As a result, exact GP inference is often impractical even for moderate numbers of computer experiments.

This computational difficulty has motivated a large literature on approximate GP methods, which reduce computational cost in various (sometimes complementary) ways. In this section, we review several popular strategies for approximate GP emulation:
\begin{itemize}
    \item \textbf{Inducing-point methods}, which approximate the latent GP using a smaller set of representative locations \citep{quinonero2005unifying, snelson2006sparse, titsias2009variational, hensman2013gaussian}. These methods are popular in the machine learning and spatial statistics communities, where extremely large datasets are common and inducing-point approaches can offer substantial computational savings. However, for high-dimensional input spaces, a large number of inducing points may be required to fill the space and adequately capture the underlying covariance structure.
    \item \textbf{Low-rank and feature approximations}, which replace the full covariance with a lower-rank representation \citep{rahimi2007random, banerjee2008gaussian, cressie2008fixedrank, finley2009improving, katzfuss2017multiresolution}. These methods can provide substantial computational savings while retaining high accuracy when the covariance structure can be adequately represented using relatively few basis functions or features, but may struggle with complex processes exhibiting high effective dimensionality.
    \item \textbf{Local, subset-based, and divide-and-conquer methods}, which reduce computational cost by replacing the global GP with a collection of smaller GP models \citep{tresp2000bayesian, keerthi2005matching, gramacy2008bayesian, gramacy2015local, rumsey2023leap}. These approaches can naturally handle nonstationarity, making them an excellent choice when modeling heterogeneous functions whose behavior changes across the input space. While they can be computationally efficient, maximizing their benefits may require additional resources, such as parallel computing or substantial memory. By relying primarily on local information, they may also lose information about long-range dependence.
    \item \textbf{Nearest-neighbor and Vecchia-type approximations}, which replace the full joint dependence structure with a sparse conditional representation \citep{vecchia1988estimation, stein2004approximating, datta2016hierarchical, katzfuss2021vecchia, katzfuss2022scaled}. These methods have emerged as particularly promising approaches to scalable GP modeling, offering substantial computational savings while often retaining excellent predictive accuracy \citep{rumsey2025all}. They can be sensitive to the choice of conditioning sets, however, and may struggle with high-dimensional inputs or capturing long-range dependence.
    \item \textbf{Structured kernel methods}, which exploit algebraic structure in the covariance matrix, such as Toeplitz or Kronecker structure, to accelerate matrix computations \citep{wilson2015kernel, gardner2018gpytorch}. These approaches can provide substantial speedups while retaining high accuracy, and they can often be leveraged in conjunction with other approaches on this list. However, they typically require specific assumptions about the covariance structure or input design.
    \item \textbf{Fast parameter estimation}, which reduces the cost of learning covariance or hyperparameters using various approximations to likelihood-based estimates \citep{rasmussen2006gpml, nickisch2008approximations, opper2009variational, gerber2021fast}. Some of these approaches can recover near-full-GP performance with a relatively small amortized cost, but they may require an expensive training phase. Because they focus on accelerating parameter estimation, the memory requirements and computational cost of prediction may remain substantial.
\end{itemize}

Each of these strategies trades exactness for scalability in a different way, and no single method is uniformly best. The most appropriate approximation depends on factors such as training set size, input dimension, characteristics of the underlying process, and available computational resources. For example, local methods may be the right choice for highly nonstationary functions, while inducing point or low-rank methods may be effective when the underlying covariance structure admits a suitable approximation. The best approach depends on the problem at hand and downstream goals, and empirical comparisons are often necessary to identify the most suitable emulator. See \cite{rumsey2025all} for an extensive empirical comparison of many of these approaches in the scalar-response setting.

We focus briefly on inducing-point methods, which are widely used and serve as the basis for one of the fully Bayesian GP approaches considered in our analysis demonstration. These methods introduce a smaller set of auxiliary inputs $\bm z_1,\dots,\bm z_m \in \mathcal{X}$, with $m \ll n$, and corresponding latent values

\begin{equation}
\bm u = \left(y(\bm z_1), \dots, y(\bm z_m)\right)^\top,
\end{equation}

which serve as a low-dimensional summary of the latent process. Let $K_n$ denote the covariance matrix over the training inputs, $K_m$ the covariance matrix over the inducing points, and $K_{nm}$ the cross-covariance between them. The resulting approximation is based on the low-rank matrix

\begin{equation}
Q_n = K_{nm}K_m^{-1}K_{mn},
\end{equation}

which replaces the full covariance structure by one mediated through the inducing points. Because computations can then be carried out largely in terms of the smaller $m \times m$ matrix $K_m$ rather than the full $n \times n$ matrix $K_n$, the computational burden is reduced substantially. Different inducing-point methods vary in how they treat the residual covariance $K_n - Q_n$, with variational formulations \citep{titsias2009variational} being especially common in modern implementations.

A number of software tools implement approximate GP methods. In Python, \texttt{GPyTorch} \citep{gardner2018gpytorch} and \texttt{GPflow} \citep{matthews2017gpflow} provide scalable GP functionality, especially for inducing-point, structured kernel, and variational methods. In \textsf{R}, packages such as \texttt{laGP} \citep{gramacy2016lagp} implement local GP approximations, \texttt{GpGp} \citep{guinness2019gpgp} provides Vecchia-type scalable GP approximations, \texttt{hetGP} \citep{binois2018hetgp} supports efficient approximate GP modeling, and \texttt{gplite} \citep{gplite} provides variational approximations for GP regression, including inducing-point GPs with variational inference.

\subsection{Basis Function Regression}

Basis function regression methods provide an alternative and widely used class of surrogate models for computer experiments. In contrast to GPs, which define the surrogate through a covariance kernel over the input space, basis function approaches represent the response using an explicit expansion in nonlinear functions of the inputs. Here, we use the term ``basis function regression'' broadly to describe any regression model in which the surrogate approximation $\hat{y}(\bm{x})$ of the simulator response $y(\bm{x})$ is represented as a finite linear expansion:
\begin{equation}\label{eq:basis_function_regression}
    \hat{y}(\bm{x}) = \sum_{m=1}^M \beta_m b_m(\bm{x}),
\end{equation}
where each $b_m: \mathcal{X} \to \mathbb{R}$ is a basis function, and $\beta_m \in \mathbb{R}$ is the associated coefficient. This representation can make basis function methods computationally attractive for large training sets, particularly when the basis has moderate dimension or can be selected adaptively \citep{rumsey2025all}.

The choice of expansion governs both approximation accuracy and UQ. A flexible basis can capture the dominant structure of the simulator response, while an inadequate one can lead to poor surrogate predictions. Unlike GP models, which induce uncertainty through a covariance function, basis function methods typically represent predictive uncertainty through probabilistic assumptions on the coefficients, selected basis terms, and residual variance. Many implementations assume a constant residual variance function $v(\bm{x})=v$, although heteroskedastic extensions allow predictive variance to vary over $\mathcal{X}$.

In the following subsections, we describe several basis function methods, including Polynomial Chaos Expansions (PCE), Multivariate Adaptive Regression Splines (MARS), Projection Pursuit Regression (PPR), and Bayesian Additive Regression Trees (BART). Although these methods differ substantially in their construction, each can be viewed as a particular realization of the general basis expansion framework described above. Not all of these methods are inherently Bayesian in their standard form. In these cases, we first introduce them generically and then describe how they are typically implemented within a Bayesian framework to propagate uncertainty in coefficients, basis selection, and model structure.

\subsubsection{Polynomial Chaos Expansions}

PCE models approximate a simulator response using polynomial basis functions \citep{wiener1938homogeneouschaos, xiu2002wiener}. In this framework, the input $\bm{x}$ is treated as a random vector with joint distribution on $\mathcal{X}$, and the response is represented using the general form in Equation~\eqref{eq:basis_function_regression}. Each basis function is constructed as a tensor product of univariate polynomials,
\begin{equation}
    b_m(\bm x) = \prod_{j=1}^p \psi^{(j)}_{\alpha_{mj}}(x_j),
\end{equation}
where $\psi^{(j)}_{\alpha_{mj}}$ denotes the degree-$\alpha_{mj}$ polynomial associated with input $x_j$. For each input dimension $j$, the polynomial family $\{\psi^{(j)}_k\}_{k \ge 0}$ is chosen to be orthogonal with respect to the marginal distribution of $x_j$, meaning that
\begin{equation}
    \int \psi^{(j)}_k(x)\,\psi^{(j)}_\ell(x)\,f_j(x)\,dx = 0, \qquad k \neq \ell,
\end{equation}
where $f_j$ is the marginal PDF of $x_j$. If the components of $\bm x$ are independent, the resulting multivariate basis functions are orthogonal with respect to the joint input distribution. Classical examples include Hermite polynomials for Gaussian inputs and Legendre polynomials for uniformly distributed inputs. For nonstandard input distributions, transformations are often used to obtain variables with suitable orthogonal polynomial bases. This orthogonal structure is useful for tasks such as sensitivity analysis and uncertainty propagation \citep{xiu2002wiener, sudret2008global, luthen2021sparsepce}.

PCE models can be computationally efficient for moderate input dimensions and polynomial degrees \citep{sudret2008global, luthen2021sparsepce}. However, their global polynomial form is best suited to smooth responses and can yield unstable or inaccurate extrapolations outside the observed input range. Responses with local irregularities, discontinuities, sharp regime changes, or complex interactions may require many terms, reducing scalability as dimension or polynomial degree increases \citep{sudret2008global, blatman2011adaptive, luthen2021sparsepce}. To control this growth, the total degree is typically restricted so that $\sum_{j=1}^p \alpha_{mj} \le d$ for some prescribed maximum $d$ \citep{sudret2008global, blatman2011adaptive}. Additional truncation strategies are often used to further reduce the basis size, especially in moderate- to high-dimensional settings \citep{blatman2011adaptive, luthen2021sparsepce}.

In a Bayesian formulation, prior distributions are placed on the coefficients $\bm \beta$, the residual variance $v$, and quantities controlling truncation or sparsity of the expansion \citep{sargsyan2014dimensionality, shao2017bayesiansparsepce, rumseyBayesianPCE}. For example, sparse basis expansions can be constructed using sequential basis-selection procedures based on information criteria \citep{shao2017bayesiansparsepce}, while shrinkage priors can be placed directly on the coefficients to favor parsimonious representations \citep{burkner2023fully}. Reversible jump MCMC methods \citep{green1995reversible} provide a fully Bayesian alternative for exploring the space of sparse expansions \citep{rumseyBayesianPCE}. Posterior inference then propagates uncertainty in the selected basis functions, polynomial coefficients $\beta_m$, and residual variance $v$ into the predictive distribution. In the demonstration analysis in Section~\ref{sec:demo}, we implement PCE using the \texttt{khaos} \textsf{R} package, which provides implementations of several of these basis-selection and sparsity methods \citep{khaos}.

\subsubsection{Multivariate Adaptive Regression Splines}

Multivariate Adaptive Regression Splines (MARS) is a flexible basis function regression method that approximates the simulator response using piecewise low-order spline functions of the inputs \citep{friedman91}. In surrogate modeling for computer experiments, MARS is attractive because it can capture nonlinear effects and interactions while remaining computationally efficient. Unlike PCE, which relies on a fixed global polynomial basis, MARS constructs its basis adaptively from the observed data.

The basic building blocks of a MARS model are hinge functions of the form
\begin{equation}
    (s(x_j - t))_+ = \max\{0, s(x_j - t)\},
\end{equation}
where $s \in \{-1,1\}$ is a sign parameter and $t \in \mathbb{R}$ is a knot location. These functions are zero on one side of the knot and linear on the other, allowing the surrogate to adapt to localized changes in the response surface.

MARS represents interactions among inputs by multiplying hinge functions. A general MARS surrogate may be written in the form of Equation~\eqref{eq:basis_function_regression}, where each basis function $b_m(\bm x)$ is a product of hinge functions involving one or more components of $\bm x$:
\begin{equation}
    b_m(\bm x) = \prod_{\ell=1}^{L_m} \left(s_{m\ell}(x_{j_{m\ell}} - t_{m\ell})\right)_+,
\end{equation}
where $L_m$ denotes the interaction order of the $m$th basis term.

This localized, piecewise structure is effective in many settings, but it can be less smooth than desired when the underlying simulator response is highly differentiable. In addition, practical implementations typically restrict the interaction order for computational reasons, which may limit performance for highly complex simulators with strong higher-order interactions.

In Bayesian implementations, prior distributions are placed on all uncertain model components, including the number of basis functions $M$, interaction orders $L_m$, active variables $x_{j_{m\ell}}$, signs $s_{m\ell}$, knot locations $t_{m\ell}$, basis coefficients $\beta_m$, and residual variance $v$ \citep{denison1998bayesianmars, francom2018bass}. Posterior inference then propagates uncertainty in these quantities into the posterior predictive distribution. The \texttt{BASS} package \citep{francom2019JSS} is used in Section~\ref{sec:demo} to demonstrate Bayesian MARS.

\subsubsection{Projection Pursuit Regression}

Projection Pursuit Regression (PPR) models a simulator response as a sum of nonlinear functions of linear projections of the inputs \citep{friedman1981ppr}. In the form of Equation~\eqref{eq:basis_function_regression}, each basis term is

\begin{equation}
    b_m(\bm x) = g_m\left(\bm \theta_m^\top \bm x\right),
\end{equation}
where $\bm \theta_m \in \mathbb{R}^p$ is a projection direction and $g_m : \mathbb{R} \to \mathbb{R}$ is a univariate nonlinear ridge function. Various forms of the ridge functions have been proposed, including nonparametric smoother functions \citep{friedman1981ppr}, GPs \citep{chen2022projectionpursuitgaussianprocess}, and spline expansions \citep{collins2024bayesian}.

PPR is especially effective when the simulator response varies primarily along a small number of important directions, even if the input dimension is moderately large, and it can capture complex interaction effects indirectly through nonlinear transformations of the projection. The main challenge of PPR is that both the projection directions and the ridge functions must be learned from the data, leading to a nonconvex and potentially multimodal inference problem. If the response does not admit a useful low-dimensional projection structure, many projection terms may be required, reducing the computational advantages.

A Bayesian formulation of PPR places prior distributions on the coefficients $\beta_m$, the projection vectors $\bm \theta_m$, the ridge functions $g_m$, the residual variance $v$, and sometimes on the number $M$ of ridge functions \citep{collins2024bayesian}. Projection directions are typically constrained to the unit hypersphere for identifiability reasons, for example using the Fisher-von Mises distribution \citep{fisher1953dispersion} or the Power Spherical distribution of \citet{decao2020powersphericaldistribution} as a prior, though this is not strictly necessary for predictive purposes. To reduce computational burden and mitigate overfitting, sparsity-inducing priors may also be placed on the projection directions, effectively constraining the surrogate to low-order interactions. Posterior inference then propagates uncertainty in both the directions $\bm \theta_m$ and the ridge functions $g_m$ into predictive uncertainty for the surrogate. In Section~\ref{sec:demo}, we use the \texttt{BayesPPR} package \citep{bayesppr} in \textsf{R} to demonstrate a fully Bayesian implementation of PPR.

\subsubsection{Bayesian Additive Regression Trees}

Bayesian Additive Regression Trees (BART; \citealp{chipman2010bart}) is a flexible nonparametric regression method that represents the simulator response as a sum of regression trees. It can adapt automatically to nonlinearities and interactions without requiring them to be specified in advance \citep{chipman2012sequential, pratola2016bartcalibration, JMLR:v26:24-1584, horiguchi2025estimatingshapleyeffectsbigdata, rumsey2025all}. We begin by introducing a single regression tree, then extend this construction to the full BART sum-of-trees model.

A single regression tree approximates the simulator response by recursively splitting the input space $\mathcal{X}$ according to binary rules of the form $x_j < c$. Each sequence of splitting decisions sends an input $\bm x$ to a unique terminal node, so a tree with $L$ terminal or ``leaf'' nodes induces the partition
\begin{equation}
    \mathcal{X} = \bigcup_{\ell=1}^L R_\ell,
\end{equation}
where the regions $R_1,\dots,R_L$ are mutually disjoint, and each region $R_\ell$ is assigned a scalar prediction $\mu_\ell$. The corresponding tree-based regression function is
\begin{equation}
    g(\bm x; T, \bm \mu) = \sum_{\ell=1}^L \mu_\ell \, \mathbf{1}\left(\bm x \in R_\ell\right).
\end{equation}

Here, $T$ denotes the tree structure, including the splitting variables and split locations, and $\bm \mu = (\mu_1,\dots,\mu_L)^\top$ are the terminal-node parameters. Thus, a regression tree gives a piecewise constant approximation to the simulator response.

BART builds on this idea by combining many regression trees additively to form the surrogate approximation
\begin{equation}
    \hat{y}(\bm x) = \sum_{k=1}^K g(\bm x; T_k, \bm \mu_k)
    = \sum_{k=1}^K \sum_{\ell=1}^{L_k} \mu_{k\ell} \, \mathbf{1}\{\bm x \in R_{k\ell}\},
\end{equation}
where $K$ is the number of trees, and each pair $(T_k,\bm \mu_k)$ defines terminal regions $R_{k\ell}$ with corresponding parameters $\bm \mu_k = (\mu_{k1},\dots,\mu_{kL_k})^\top$. To connect this representation with Equation~\eqref{eq:basis_function_regression}, each terminal node can be viewed as defining a basis function,
\begin{equation}
    b_m(\bm x) = \mathbf{1}\{\bm x \in R_{k\ell}\},
\end{equation}
with coefficient $\beta_m = \mu_{k\ell}$, where $m$ is a unique index assigned to the terminal node indexed by $(k,\ell)$.

Because trees partition the input space recursively, BART can adapt to localized features, discontinuities, and interaction structure in ways that are often difficult for global polynomial or spline models. BART is also relatively robust in practice, since the additive structure and prior regularization reduce the tendency of a single large tree to overfit. Its main limitation is that the surrogate is built from piecewise constant components, so the implied mean function is not smooth in the same sense as a GP or spline-based model. This can be a disadvantage when the simulator response is known to be highly smooth.

A Bayesian formulation is central to BART. Prior distributions are placed on the tree structures $T_1,\dots,T_K$, terminal-node parameters $\bm \mu_1,\dots,\bm \mu_K$, and residual variance $v$. The tree-structure prior favors shallow trees, so that each tree explains only a small part of the overall response. Posterior inference is usually carried out by Markov chain Monte Carlo, with tree structures updated through local modifications such as grow, prune, change, and swap moves, often using reversible jump MCMC \citep{green1995reversible}. As with the other basis function regression approaches considered here, the residual variance is often assumed constant over the input space, although heteroskedastic extensions have also been developed \citep{pratola2020heteroscedastic}. While several BART implementations are available, we use the \texttt{stochtree} package \citep{stochtree} for the analysis demonstration in Section~\ref{sec:demo}, as it is well suited to the surrogate modeling context and provides fast fitting and prediction capabilities.

\subsection{Surrogate Models for High-Dimensional Response}\label{sec:surr_HD}

The univariate surrogate models described above provide the foundation for scalar simulator responses, but many computer experiments produce vector-, field-, or function-valued outputs that require additional modeling structure. Surrogate modeling becomes substantially more challenging when simulator outputs $\bm y(\bm x)$ are high-dimensional, since the model must capture both input-output dependence and dependence across output components. In such settings, the output space $\mathcal{Y}$ is a subset of $\mathbb{R}^q$, where $q$ may be in the hundreds, thousands, or even larger. Broadly speaking, three primary approaches are used in this setting: (i) specifying a joint multivariate stochastic process model across outputs, (ii) fitting separate emulators for each output while borrowing strength across outputs, and (iii) performing dimension reduction on the outputs. The choice among these approaches depends on the size and structure of the output, although the dimension-reduction-based approach of \cite{higdon2008computer} is by far the most common in modern computer experiments. We briefly review these methods here.

The first strategy is to specify a joint model for the entire response, allowing dependence between outputs to be modeled directly. In the popular approach of \cite{ContiOHagan2010}, a separable covariance is defined as the product of the covariance over the simulator inputs and a covariance over the outputs. Other multivariate surrogate constructions include the linear model of coregionalization \citep{JournelHuijbregts1978} and nonseparable covariance models \citep{FrickerOakleyUrban2013}. Application-specific covariance structures have also been proposed, for example by \cite{mak2018efficient}. For ordered outputs, such as time series or spatial fields, an appealing alternative is to treat the output index as another input to the emulator, producing the ``Time-Input'' emulator of \cite{ContiOHagan2010}, originally proposed by \cite{kennedy2001bayesian}. When outputs are recorded on a common grid, this approach can be extremely efficient by exploiting the resulting tensor-product structure in the design matrices \citep{francom2019JSS}.

Perhaps the most straightforward approach is to fit a fully independent surrogate model for each output. This corresponds to the special case of the approach described by \cite{ContiOHagan2010} with a diagonal cross-output covariance matrix. \cite{FrickerOakleyUrban2013} showed that this approach can perform competitively in some situations, and the approach has been applied successfully to model responses over a temporal domain in several applications including \citet{lee2013magnitude, brown2018, mak2018efficient}. In sufficiently complex problems, however, ignoring the dependence between outputs can lead to poor prediction and UQ capabilities. \cite{gu2016parallel} improve upon this idea by fitting separate GP surrogates while jointly estimating select covariance parameters for the GP kernel across outputs, allowing the models to borrow strength across output dimensions while retaining much of the computational efficiency of the independent approach. 

Although joint and partially pooled approaches provide principled ways to model dependence across outputs, they can become computationally demanding as the number of outputs increases. A common alternative is to first project the response onto a lower-dimensional basis \citep{higdon2008computer}, writing
\begin{equation}\label{eq:pca}
\bm y(\bm x) \approx \bm\mu + \sum_{k=1}^K w_k(\bm x)\,\bm\phi_k,
\end{equation}
where $\bm\mu \in \mathbb{R}^q$ is a mean vector, $\bm\phi_1,\ldots,\bm\phi_K \in \mathbb{R}^q$ are basis vectors, and $w_1,\ldots,w_K:\mathbb{R}^p\to\mathbb{R}$ are scalar coefficient functions over the input space. The basis is often constructed from empirical principal components obtained via a singular value decomposition of the centered training-output matrix, although wavelets, splines, and other physically motivated representations are also used \citep{bayarri2007AS,higdon2008computer,Wilkinson2010}. Surrogate models are fit to the coefficient functions $w_k(\bm x)$, often independently across $k$ for computational convenience. This reduces the problem of emulating a high-dimensional output to a collection of scalar-output surrogate models and is a widely used strategy for high-dimensional simulator outputs.

In practice, the effectiveness of this dimension-reduction strategy depends not only on the choice of basis but also on the number of basis components retained in the finite approximation. For principal component decompositions, a common default is to retain enough components to explain a specified proportion of the total variance. However, later components often represent increasingly fine-scale or noisy variation and may be more difficult to emulate accurately as functions of the inputs. Including too many components can therefore degrade out-of-sample performance by adding coefficient functions that are weakly predictable and prone to overfitting. Cross-validation of the surrogate models fitted to each $w_k(\bm x)$ provides a useful diagnostic for identifying this behavior and selecting an appropriate truncation level.

Once surrogate models have been fit to the coefficient functions, predictive uncertainty in the reconstructed output arises from two distinct sources. The first is uncertainty in the predicted coefficient functions themselves. If the surrogate for the coefficient function $w_k$ has predictive variance $v_k(\bm x)$, the corresponding covariance matrix in output space is
\begin{equation}
\bm\Sigma_{\mathrm{surr}}(\bm x)
=
\bm\Phi
\operatorname{diag}\!\left\{
v_1(\bm x),\ldots,
v_K(\bm x)
\right\}
\bm\Phi^{\top},
\end{equation}
where $\bm\Phi=[\bm\phi_1,\ldots,\bm\phi_K]$. This captures uncertainty in the surrogate predictions for the retained basis coefficients.

The second source of uncertainty is approximation error due to truncating the basis expansion. To characterize this type of uncertainty, define the basis-truncation error as
\begin{equation}\label{eq:trunc_err}
\bm\epsilon_{\mathrm{trunc}}(\bm x)
=
\bm y(\bm x)
-
\left(
\bm\mu+\sum_{k=1}^K w_k(\bm x)\bm\phi_k
\right).
\end{equation}
Because only the first $K$ basis functions are retained, this error cannot be eliminated by improving the surrogate models for the coefficient functions. It must instead be characterized separately and propagated alongside uncertainty in the coefficient-function predictions.

There are two common approaches for propagating uncertainty due to truncation error. The first is to assume a multivariate Gaussian distribution,
\begin{equation}
\bm\epsilon_{\mathrm{trunc}}(\bm x)
\sim
\mathcal{N}(\bm 0,\bm\Sigma_{\mathrm{trunc}}),
\end{equation}
where $\bm\Sigma_{\mathrm{trunc}}$ is estimated from the observed truncation errors across the training inputs. The resulting predictive covariance is
\begin{equation}
\bm\Sigma_{\mathrm{pred}}(\bm x)
=
\bm\Sigma_{\mathrm{surr}}(\bm x)
+
\bm\Sigma_{\mathrm{trunc}},
\end{equation}
which serves as the covariance matrix of a multivariate Gaussian distribution representing predictive uncertainty.

An alternative approach avoids imposing a parametric distribution on the truncation error \citep{francom2019inferring}. For each training input $\bm x_i$, the truncation error $\bm\epsilon_{\mathrm{trunc}}(\bm x_i)$ can be computed using the formula in Equation~\eqref{eq:trunc_err}, and the truncation error at a new input $\bm x$ can be represented by the empirical discrete distribution
\begin{equation}\label{eq:emp_trunc}
\bm\epsilon_{\mathrm{trunc}}(\bm x)
\sim
\operatorname{Unif}
\left\{
\bm\epsilon_{\mathrm{trunc}}(\bm x_1),
\ldots,
\bm\epsilon_{\mathrm{trunc}}(\bm x_n)
\right\},
\end{equation}
from which samples can be taken to propagate uncertainty due to truncation error without imposing assumptions about the truncation-error distribution.

The choice between these two approaches depends on the amount of available training data and the degree of structure that can reasonably be assumed for the truncation error. The Gaussian approach provides a simple and computationally convenient representation, while the empirical approach offers greater flexibility while retaining computational tractability. In either case, the truncation-error model complements surrogate uncertainty in the retained coefficient functions, providing a more complete characterization of predictive uncertainty in the reconstructed output.

\section{Surrogate Model Applications}\label{sec:use_cases}

Previous sections focused on constructing surrogate models, from scalar-output emulators to methods for high-dimensional responses. We now turn to how these models are used in practice. Because surrogate models provide fast statistical approximations to expensive simulations, they play a central role in many UQ tasks where repeated model evaluations are required. In this section, we review three common applications of surrogate modeling in UQ: forward UQ, sensitivity analysis, and Bayesian model calibration.

\subsection{Forward Uncertainty Quantification}

Forward UQ is the process of characterizing how uncertainties in input parameters propagate to the outputs of a computational or mathematical model used to gain insight into a physical or engineering system. Input uncertainties can arise from various sources, including incomplete knowledge of the physical system, inherent variability, and measurement error. From a statistical perspective, the primary goal of forward UQ is to provide a probabilistic characterization of the outputs given uncertainty in the inputs, thereby supporting improved decision-making.

As described in \cite{smith2013}, the general steps in forward UQ include:

\begin{itemize}
\item Modeling input uncertainties as random variables with specified probability distributions.
\item Propagating these uncertainties through the computational or mathematical model using an experimental design that enables the construction of a computationally efficient surrogate of the original model.
\end{itemize}

After propagation, the resulting output distribution is typically summarized through application-specific quantities of interest. For high-dimensional simulator outputs, this involves characterizing the induced distribution of an entire vector-, field-, or function-valued response rather than a single scalar output. When a basis representation such as Equation~\eqref{eq:pca} is used, uncertainty in the inputs is propagated through the surrogate models for the coefficient functions $w_k(\bm x)$, and mapped back to the original output space. This allows one to estimate pointwise summaries for individual output components, joint summaries that account for dependence across components, or lower-dimensional functionals of the high-dimensional output, such as spatial averages, extrema, or exceedance probabilities.

These summaries may be complemented by exploratory data analysis to characterize the statistical properties of the model outputs, as well as by sensitivity analysis to determine which input uncertainties have the greatest effect on output variability. Sensitivity analysis can also reveal the nature of the relationships between inputs and outputs, such as whether they are linear, nonlinear, or dominated by interactions. A more technical description of sensitivity analysis is presented in the next subsection.

\subsection{Sensitivity Analysis}

Sensitivity analysis is used to quantify how uncertainty in model inputs contributes to variability in the model output. Broadly speaking, sensitivity analysis methods can be classified as either global or local. Global methods assess the influence of inputs across the full range of their distribution, while local methods characterize the effect of small perturbations around a given input setting. In the context of UQ, global methods are often more relevant because they characterize main effects and interactions across the entire input space, yielding an overall summary of variable importance relative to the response of interest.

A widely used global sensitivity analysis framework is the variance-based approach of Sobol' \citep{Sobol2001}, with further discussion in \citet{saltelli2007}. Writing $\bm{x} = \left(x_1,\dots,x_p\right)$, the Sobol' decomposition expresses the model response $y(\bm{x})$ as a sum of component functions corresponding to main effects, second-order interactions, and higher-order interactions up to the full $p$-way interaction:

\begin{align}
y(\bm{x}) = y_0 + \sum_{i=1}^p y_i(x_i) + \sum_{i=1}^p \sum_{j>i} y_{ij}(x_i, x_j) + \ldots + y_{1\ldots p}(x_1, x_2, \ldots, x_p).
\end{align}

Each term in this sum other than $y_0$ has zero mean and is orthogonal to all the other terms. As shown in \cite{francom2019JSS}, these terms can be constructed using centered conditional expectations. Squaring the decomposition and taking expectations eliminates all cross-product terms by orthogonality, leaving only the squared contribution of each component and yielding the variance decomposition

\begin{align}
Var(y(\bm x))= \sum_{i=1}^p Var \left(y_i(x_i) \right) + \sum_{i=1}^p \sum_{j > i} Var\left(y_{ij}(x_i,x_j) \right) + \ldots + Var\left(y_{1\ldots p}(x_1,x_2,\ldots,x_p) \right).
\end{align}

This yields a decomposition of the output variance into the sum of variances due to each main effect, each two-way interaction after accounting for main effects, and so forth. Relative sensitivity indices for main effects and interactions are then defined as proportions of the total variance.

For high-dimensional response variables, Sobol' indices are often applied after reducing the response to scalar quantities. In basis representations such as \eqref{eq:pca}, this can be done by computing sensitivity indices for the coefficient functions $w_k(\bm x)$, thereby identifying which inputs drive variation in the dominant modes of the response. These coefficient-level indices can be examined individually, aggregated across basis terms, or propagated through the basis expansion to produce component-wise or location-specific sensitivity summaries for the original response $\bm y(\bm x)$.

For a given surrogate model $\hat{y}(\bm x)$, this variance decomposition can be computed across MCMC samples of the fitted model to obtain posterior distributions of sensitivity indices. For some surrogate classes, the variance decomposition can be computed analytically from the fitted model. Perhaps the best-known example is polynomial chaos expansions, where the orthogonality of the basis yields closed-form Sobol' indices directly from the coefficients \citep{sudret2008global}. Similar closed-form calculations are available for MARS \citep{francom2018bass, francom2019JSS} and BART \citep{horiguchi2021assessing} models. In other settings, however, analytical calculations are not feasible and Monte Carlo approximations are required.

\subsection{Bayesian Model Calibration}

Bayesian model calibration is a framework for learning uncertain model parameters by combining computer model output with experimental or observational data. Unlike optimization-based approaches, it provides a coherent way to quantify uncertainty in the calibrated parameters while accounting for measurement error and possible discrepancy between the model and reality. We begin by describing methods for calibrating models with a univariate response, then review standard extensions for high-dimensional and functional data.

\subsubsection{Calibrating to Univariate Response}

The traditional Bayesian approach to model calibration, introduced by \cite{kennedy2001bayesian} and extended by \cite{higdon2004combining}, uses physical observations to infer uncertain model parameters. Let $y(\bm{x}, \bm{u})$ denote computer model output from computational experiments, where $\bm{x}$ represents known inputs, such as experimental settings, and $\bm{u}$ denotes uncertain parameters to be calibrated. Let $z(\bm{x})$ denote the corresponding observation from the physical system. In some applications, $\bm{x}$ may also include indices such as space or time, though these can instead be handled using the methods for high-dimensional or functional responses described in subsequent subsections. If observations are collected at $n_z$ physical input settings $\bm{x}_1, \dots, \bm{x}_{n_z}$, the calibration model is

\begin{align}
z(\bm x_i) &= y(\bm x_i,\bm \theta) + \delta(\bm x_i) + \epsilon_z(\bm x_i), 
\qquad \epsilon_z(\bm x_i) \overset{\text{i.i.d.}}\sim \mathcal{N}(0,\sigma_z^2),
\label{eq:koh1}
\end{align}
where $\bm\theta$ denotes the true or best-fitting value of the calibration parameter $\bm u$, $\delta$ denotes model discrepancy between the simulator and physical system, and $\epsilon_z$ is measurement error in $z$. The likelihood specified here assumes independent and identically distributed measurement errors conditional on $\bm x_i$, though other error structures can be used. The unknowns in this model are the calibration parameters $\bm \theta$, the measurement error variance $\sigma^2_z$, and the discrepancy function $\delta$. For this model to be identifiable, prior distributions for each of the unknowns need to be carefully chosen. When $n_z$ is small, the prior on $\sigma^2_z$ can be influential. Additionally, the model parameters $\bm{\theta}$ and discrepancy $\delta$ are confounded, so that at least one of these must be well-constrained by the prior. For example, one may place a GP prior on $\delta$ that favors smooth, small-magnitude functions \citep{higdon2004combining}, or impose structure such as positivity or monotonicity \citep{brynjarsdottir2014learning}.

When the computer model $y$ is expensive to evaluate, repeated likelihood evaluation is often made feasible by replacing $y$ with a surrogate trained on a limited set of simulator runs \citep{sacks1989SS}. A full Bayesian analysis conditions on both simulator runs and physical observations, jointly inferring the calibration parameters, discrepancy, measurement error, and surrogate model parameters. \cite{higdon2004combining} adopt this approach using a full GP surrogate, whereas \cite{kennedy2001bayesian} fix some GP parameters in advance for computational and identifiability reasons. For both philosophical and computational reasons, many modern analyses instead adopt modularized approaches, in which the surrogate is first fit to the computational experiments and then treated as fixed during calibration. Such modularization can also improve inference by limiting feedback between model components \citep{liu2009modularization}.

\subsubsection{Calibrating to High-Dimensional Response}

Calibration becomes substantially more challenging when the response is high-dimensional. In this setting, each observation at input $\bm x_i$ is no longer a scalar but a $q$-dimensional vector, which increases both the statistical complexity of the model and the computational cost of emulation and inference. Let the observed response, simulator output, and model discrepancy be denoted by $\bm z(\bm x_i)$, $\bm y(\bm x_i,\bm u)$, and $\bm \delta(\bm x_i)$, respectively. The calibration model can then be written as

\begin{align}
    \bm z(\bm x_i) = \bm y(\bm x_i,\bm \theta) + \bm \delta(\bm x_i) + \bm \epsilon_{\bm z}(\bm x_i), 
    \qquad 
    \bm \epsilon_{\bm z}(\bm x_i) \overset{\text{i.i.d.}}\sim N(\bm 0, \sigma_{\bm z}^2 \mathcal{I}_q),
\end{align}
where $\mathcal{I}_q$ is the $q \times q$ identity matrix and $\sigma_{\bm z}^2$ represents the observational error variance.

A common strategy is to reduce the response through a basis expansion and calibrate in the resulting coefficient space. When emulation is needed, independent surrogate models are typically fit to the basis coefficients, as described in Section~\ref{sec:surr_HD}. For example, \citet{higdon2008computer} use principal components for the simulator output and a separate flexible basis for discrepancy, whereas \citet{bayarri2007framework} use a wavelet basis for both. \citet{francom2019inferring} take a different approach, projecting the simulator output onto principal components while leaving discrepancy in the original high-dimensional space, with surrogate fitting handled in a modular framework.

\subsubsection{Elastic Functional Bayesian Model Calibration} \label{sec:efbmc}

When the high-dimensional response is functional in nature, specialized calibration methods can be used. Let $t$ denote the functional variable (e.g., time), typically observed on a discrete grid, and let $z(t,\bm x_i)$ denote the experimental measurement from the $i$th experiment at functional index $t$. Similarly, let $y(t,\bm x_i,\bm u)$ denote the corresponding simulated response at $t$ under experimental conditions $\bm x_i$ and input parameters $\bm u$. A functional Bayesian calibration model takes the form
\begin{align}
    z(t,\bm x_i) &= y(t,\bm x_i,\bm \theta) + \delta(t, \bm x_i) + \epsilon_z(t,\bm x_i), ~~\epsilon_z(t,\bm x_i)\overset{\text{i.i.d.}}\sim \mathcal{N}(0,\sigma_z^2). \label{eq:fbmc}
\end{align}

Several methods extend Bayesian calibration to functional responses \citep{bayarri2007framework, higdon2008computer, brown2018}. As with other high-dimensional responses, a common approach is to represent the functions with a basis decomposition and perform calibration in the resulting coefficient space. These methods are most effective when important features of the response are aligned across $t$. When responses are misaligned, standard basis representations and distance metrics can be misleading, since they tend to capture only amplitude variation while neglecting variation in the locations of features along $t$, commonly called phase variability.

Elastic functional Bayesian model calibration (EFBMC) extends this framework to account more directly for both amplitude and phase variability. Using elastic functional data analysis (EFDA), one can define a proper distance between $z(t,\bm x_i)$ and $y(t,\bm x_i,\bm \theta)$ that accounts for misalignment \citep{srivastava2016}. This distance induces a warping function $\gamma_{y\rightarrow z}(t): [0,1] \rightarrow [0,1]$ that aligns $y(t,\bm x_i,\bm \theta)$ to $z(t,\bm x_i)$ in phase space. Through this alignment, each functional response is decomposed into two components: a transformed, temporally aligned function and its associated warping function. This decomposition allows the calibration procedure to address amplitude and phase variability explicitly, rather than forcing phase misalignment to be absorbed into the discrepancy term. 

The calibration model in Equation~\eqref{eq:fbmc} is then applied jointly to the aligned functions and the warping functions, which take on separate experimental indices $i$. After calibration, the inverse warping functions can be applied to the aligned functions to map posterior samples back to the original data space. In our experience, this procedure reduces modeling error and leads to improved inference. An implementation of this approach is available in \textsf{R} \citep{fdasrvf} and \textsf{Python} \citep{fdasrsf}. For additional details on Bayesian calibration and EFBMC, we refer the interested reader to \cite{francom2025juq}.

\subsubsection{Parameter Identifiability}

Parameter identifiability has long been a challenge for Bayesian model calibration since the introduction of the seminal Kennedy and O'Hagan (KOH) framework \citep{kennedy2001bayesian}. The KOH framework is still widely considered a gold standard for interpolative prediction \citep{tuo2018prediction}, but has serious drawbacks when the inverse problem (i.e., posterior inference for the calibration parameters) is the primary goal. In particular, the discrepancy function and calibration parameters are fundamentally confounded and cannot be uniquely separated without making assumptions about the form of the model error \citep{arendt2012quantification,brynjarsdottir2014learning}. Therefore, under the standard KOH framework, the prior for the discrepancy function can remain a permanent fixture of the calibration parameter posterior regardless of how much data is collected \citep{plumlee2017bayesian}.

A substantial literature has proposed strategies for addressing identifiability challenges through additional information or assumptions, including informative or constrained discrepancy models, multiple responses, regularization, alternative definitions of the calibration target, and modular Bayesian approaches. We provide a more detailed discussion of these approaches in Appendix A; see also Section 6 of \citet{sung2024review} for a comprehensive review of parameter identifiability in Bayesian model calibration.

\section{Demonstration of Methods}\label{sec:demo}

In this section, we provide a practical illustration of the methods described above using both simulated and real data. In both cases, the responses are multivariate. The simulated example has a modest response dimension of 15 and is intended as a transparent illustrative example, whereas the real responses are functional data. These examples are used to evaluate surrogate predictive performance through cross-validation and to illustrate downstream analyses, including Sobol' sensitivity analysis and Bayesian model calibration.

For both GP and basis-function surrogates, we use the common approach of first reducing the dimension of the response and then independently modeling the resulting basis coefficients. Predictive uncertainty is characterized using the Gaussian methodology described in Section~\ref{sec:surr_HD}, with the predictive covariance \(\bm\Sigma_\text{pred}(\bm x) = \bm\Sigma_\text{surr}(\bm x) + \bm\Sigma_\text{trunc}\) accounting for uncertainty from both the surrogate models and basis truncation. All surrogates are fit using the \texttt{mvBayes} package in \textsf{R} \citep{mvBayes}, which automates the dimension-reduction, surrogate-modeling, and cross-validation steps.

Because surrogate performance can vary substantially depending on the nature of the data \citep[see, e.g.,][]{rumsey2025all}, it is important in practice to compare multiple candidate models. The comparisons presented below should therefore be viewed as illustrative of the modeling and assessment process, rather than as definitive rankings of the methods. The purpose of this section is not to identify a universally preferable method or to provide an exhaustive analysis of these datasets, but rather to demonstrate a practical workflow and evaluation metrics for the methods reviewed here.

\subsection{Example Simulation Study}\label{sec:sim_example}

In this section, we illustrate several of the surrogate modeling approaches outlined in this review, for a synthetic multivariate-output problem designed to mimic features commonly encountered in computer experiments. The intent is to demonstrate an assessment of predictive performance and practical suitability of each surrogate for downstream UQ tasks, including sensitivity analysis and Bayesian model calibration, using a concrete exemplar dataset.

The data are simulated as follows. We generate inputs independently from a uniform distribution on the unit hypercube,
\begin{align}
    \bm x_i \overset{\mathrm{iid}}{\sim} \mathrm{Unif}([0,1]^{12}), \qquad i=1,\dots,n,
\end{align}
with $n=500$. Each generated input is then transformed as needed to match the native input domain of the underlying benchmark functions used in the simulation study.

The latent simulator is a vector-valued function $f:\mathbb{R}^{12}\to\mathbb{R}^{6}$, whose six independent components are given by scaled versions of the standard test functions ``ignition,'' ``wingweight,'' ``borehole,'' ``piston,'' ``steel column,'' and ``robot,'' all available in \cite{surjanovic_bingham_testfunctions}, with easy access in the \texttt{duqling} package in \textsf{R} \citep{duqling}. The scaling of each test function is chosen to yield approximately equal variance across components. Writing
\begin{align}
    f(\bm x) = \left(f_1(\bm x),\dots,f_6(\bm x)\right)^\top,
\end{align}
each component function $f_j$ depends only on a subset of $p_j<12$ of the 12 inputs, with the remaining $12-p_j$ inputs being inert for that component.

To induce dependence among outputs and create a higher-dimensional response, we apply a linear transformation to the latent six-dimensional outputs. Let $\Phi \in \mathbb{R}^{15\times 6}$ be a fixed transformation matrix constructed so that each row contains exactly two ones and four zeros, corresponding to a distinct pair of components of $f(\bm x)$. Thus, the transformed mean response is
\begin{align}
    \bm \mu(\bm x) = \Phi f(\bm x),
\end{align}
which is 15-dimensional. Equivalently, each component of $\bm\mu(\bm x)$ is the sum of a unique pair of latent outputs.

The observed response is then generated by adding Gaussian noise:
\begin{align}
    \bm y_i = \Phi f(\bm x_i) + \bm \epsilon_i, \qquad \bm \epsilon_i \overset{\mathrm{iid}}{\sim} \mathcal{N}(\bm 0, 0.04\,\mathcal{I}_{15}),
\end{align}
for $i=1,\dots,n$, where $\mathcal{I}_{15}$ denotes the $15\times 15$ identity matrix. This construction provides a useful exemplar problem by featuring varying degrees of nonlinearity in the latent components $f_j$, inert inputs to demonstrate robustness against false discoveries, and dependencies across $\bm \mu$ that make it a genuinely multivariate problem.

We demonstrate a comparison across the following surrogate models: a full GP where hyperparameters are sampled from their posterior distributions using a custom adaptive Metropolis sampler (``gp'' in the figures), a full GP fit using variational inference in the \texttt{gplite} package in \textsf{R} (``gplite'' in the figures; \citealp{gplite}); an inducing-point GP with variational inference, also implemented in \texttt{gplite} (``fitcgp''); Bayesian projection pursuit regression fit using the \texttt{BayesPPR} package (``bppr''; \citealp{bayesppr}); Bayesian MARS using the \texttt{BASS} package (``bass''; \citealp{bass}); BART using the \texttt{stochtree} package (``stochtree''; \citealp{stochtree}); and a Bayesian PCE with adaptive polynomial bases using \texttt{khaos} (``khaos''; \citealp{khaos}). We also compare against a manually implemented Bayesian linear model (BLM) as a baseline method (``blm''); while complex physical experiments seldom feature purely linear relationships, and while we expect the BLM to be a poor surrogate in most scenarios, including this simple baseline allows us to quantify the difference from baseline and serves as a diagnostic check for overfitting by the more flexible methods. Convergence of all algorithms is evaluated in Appendix B.

\subsubsection{Cross-Validation}

We now demonstrate a standard $K$-fold cross-validation with $K=5$, in which the $n=500$ observations are randomly partitioned into five folds of size 100. For each fold, the surrogate is trained on the remaining 400 observations and evaluated on the held-out fold. This procedure is repeated for each surrogate model under consideration.

\begin{figure}[!htbp]
    \centering
    \includegraphics[width=1.0\textwidth]{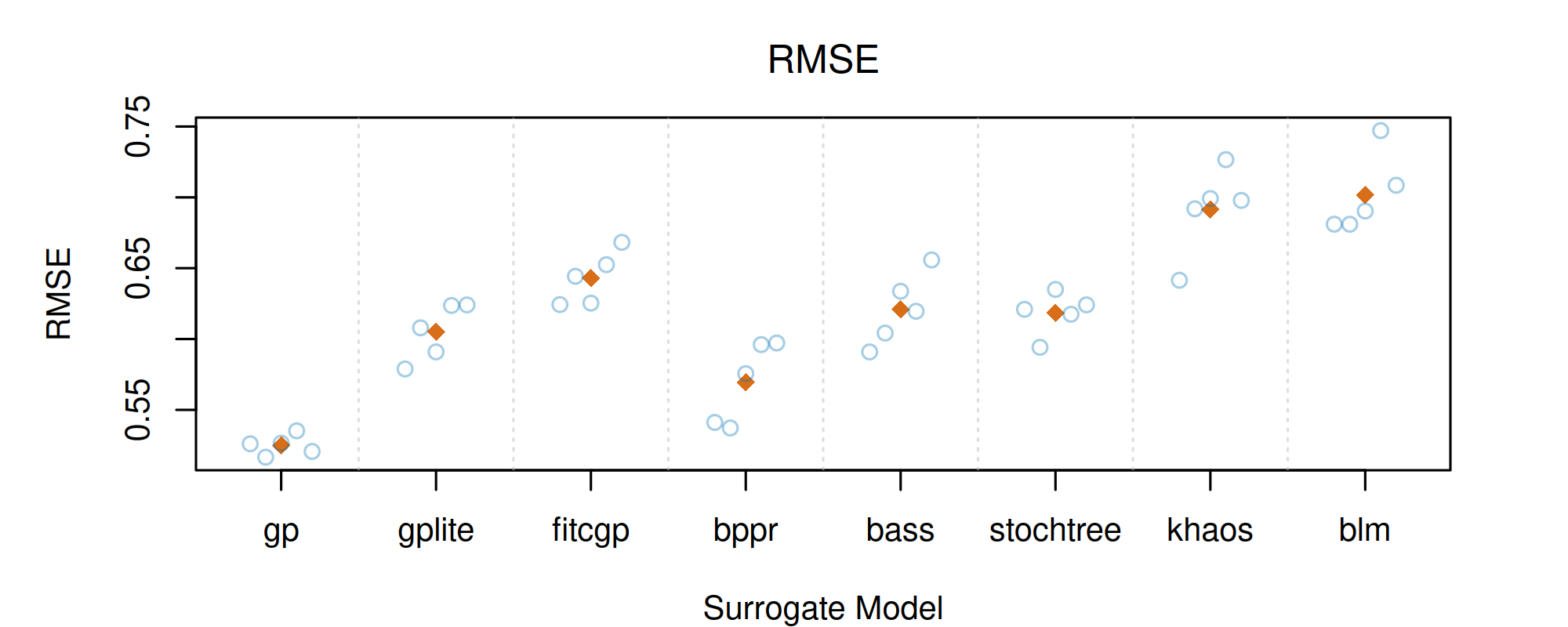} 
    \caption{Root mean squared error (RMSE) on the held-out data across the five cross-validation folds for each surrogate model.}
    \label{fig:RMSE}
\end{figure}

Performance is assessed using metrics that reflect both predictive accuracy and UQ quality. Prediction accuracy is summarized using the average root mean squared error (RMSE) across output variables, computed over the held-out responses. Figure~\ref{fig:RMSE} displays the RMSE for the held-out set in each of the five folds, with open circles indicating results per fold, ordered from left to right by fold, and orange diamonds indicating mean RMSE across folds for each method. The full GP has the lowest RMSE in each fold, demonstrating the best predictive accuracy in this example. BayesPPR is the next strongest performer, with GPLite, BASS, and Stochtree not far behind. Khaos has accuracy similar to BLM for this example, likely indicating overfitting.

\begin{figure}[!htbp]
    \centering
    \includegraphics[width=1.0\textwidth]{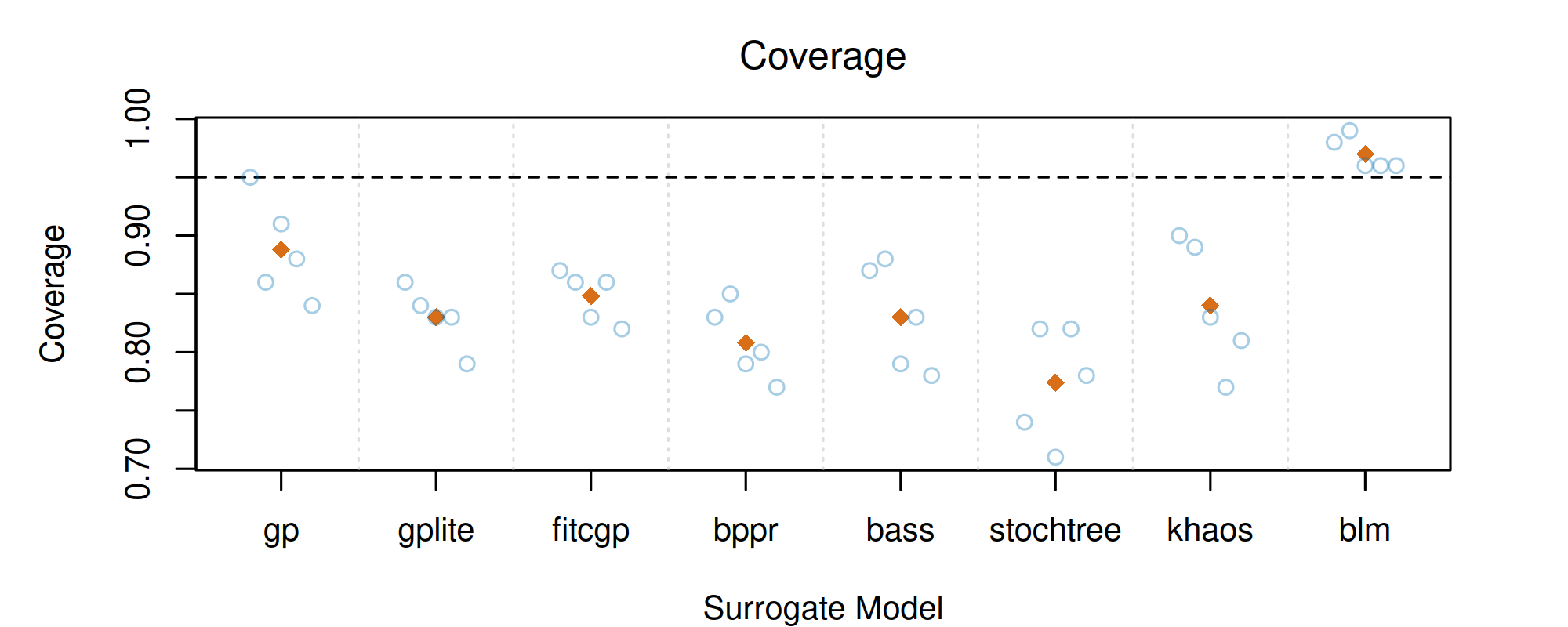} 
    \caption{Empirical coverage of posterior predictive intervals across the five cross-validation folds for each surrogate model. Values closer to the nominal level indicate better calibrated predictive uncertainty.}
    \label{fig:cover}
\end{figure}

To evaluate predictive uncertainty, we first examine empirical coverage of posterior predictive intervals. For a multivariate response, we compute a prediction region using the extremal depth method, as outlined in \cite{narisetty2016extremal}, which results in a $1-\alpha$ simultaneous prediction region. Coverage is then assessed by recording the proportion of held-out responses that lie entirely inside this region, that is, within the region in all components simultaneously (here $1-\alpha=0.95$). Figure~\ref{fig:cover} shows the resulting empirical coverage. All methods, aside from the baseline BLM, tend to under-cover to some degree. The full GP is closest to attaining nominal coverage, and Stochtree is the furthest from doing so. Although no method is perfectly calibrated, all provide usable predictive UQ, as demonstrated by the downstream analyses in the next subsection.

\begin{figure}[!htbp]
    \centering
    \includegraphics[width=1.0\textwidth]{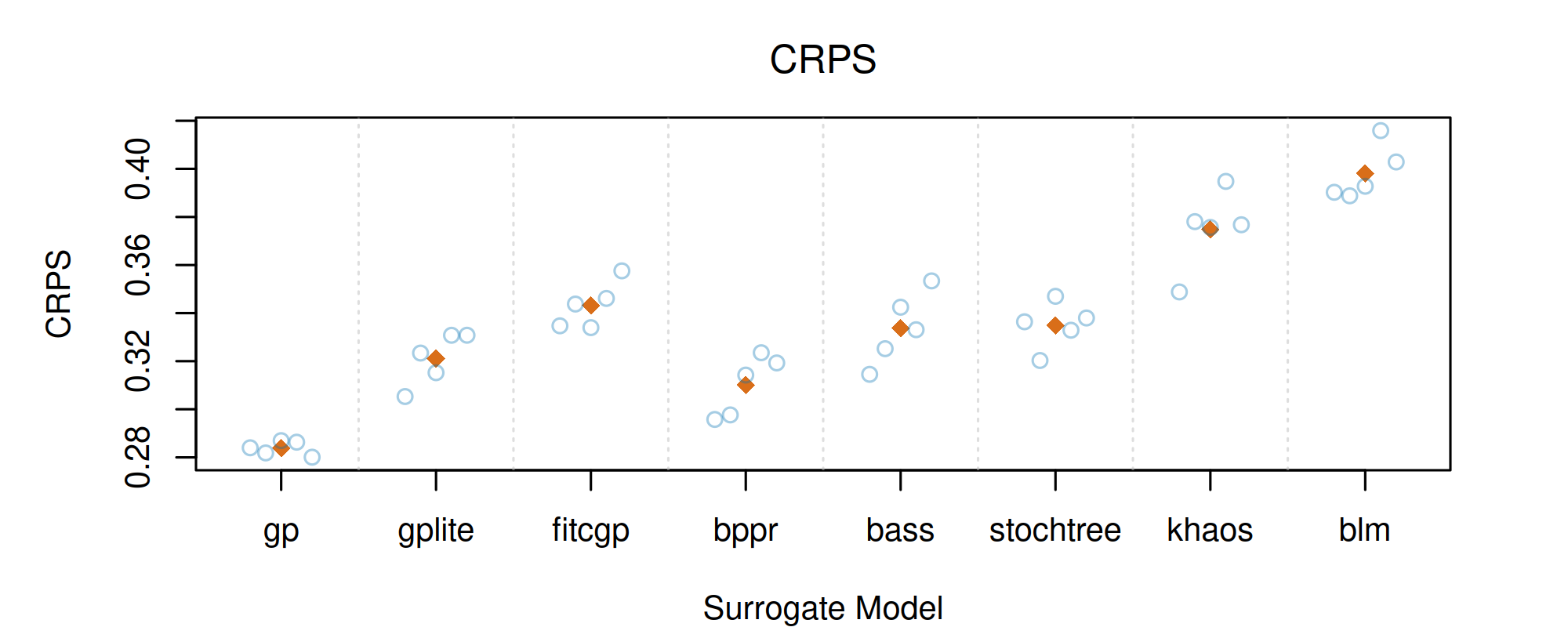}
    \caption{Continuous ranked probability score (CRPS) across the five cross-validation folds for each surrogate model. Lower values indicate better predictive UQ.}
    \label{fig:crps}
\end{figure}

To assess UQ in a more holistic way, we also evaluate the average marginal continuous ranked probability score (CRPS) of the posterior predictive distributions. CRPS is a proper scoring rule for probabilistic forecasts that rewards narrow distributions centered around the truth while penalizing biases and overly dispersed distributions \citep{gneiting2007strictly}. Lower values indicate better performance. Figure~\ref{fig:crps} shows that the full GP has the lowest CRPS, followed by BayesPPR and GPLite. As with RMSE, the CRPS of the Khaos surrogate is only slightly lower than that of the BLM.

We also evaluate computational cost, including both model fitting and prediction time. In downstream tasks such as global sensitivity analysis, uncertainty propagation, and Bayesian calibration, even modest differences in computational efficiency can become consequential because of the need for repeated evaluation of the surrogate model. Table~\ref{tab:timing_comparison} reports the median per-fold fit and prediction times for each method. As expected, the full GP is by far the slowest for both fitting and prediction, and no method comes close to the fitting speed of the BLM with its conjugate posterior distributions, although GPLite and FitCGP have comparable prediction times. Among the remaining methods, GPLite and FitCGP are the fastest to fit and predict, and Khaos is the slowest, with BayesPPR, BASS, and Stochtree falling in the intermediate range. We note that relative timing will vary according to data size and complexity.
\begin{table}[!htbp]
    \centering
    \caption{Median fit and prediction times, in seconds, for each surrogate model.}
    \label{tab:timing_comparison}
    \resizebox{\textwidth}{!}{%
    \begin{tabular}{lrrrrrrrr}
        \hline
        Method & GP & GPLite & FitCGP & BayesPPR & BASS & Stochtree & Khaos & BLM \\
        \hline
        Fit Time (s)     & 458.72 & 5.06 & 4.49 & 26.13 & 20.63 & 38.17 & 95.54 & 0.01 \\
        Predict Time (s) & 44.90 & 0.21 & 0.20 & 0.48 & 0.27 & 1.34 & 3.84 & 0.17 \\
        \hline
    \end{tabular}
    }
\end{table}

These results illustrate how cross-validation can be used to compare surrogate models for a particular dataset, rather than suggesting general conclusions about their relative performance. Results may vary with fold selection and performance depends strongly on the dataset. For a broader evaluation of method timing and performance across datasets of various sizes, see \cite{rumsey2025all}.

\subsubsection{Downstream Analyses}

\begin{figure}[!htbp]
    \centering
    \includegraphics[width=0.9\textwidth]{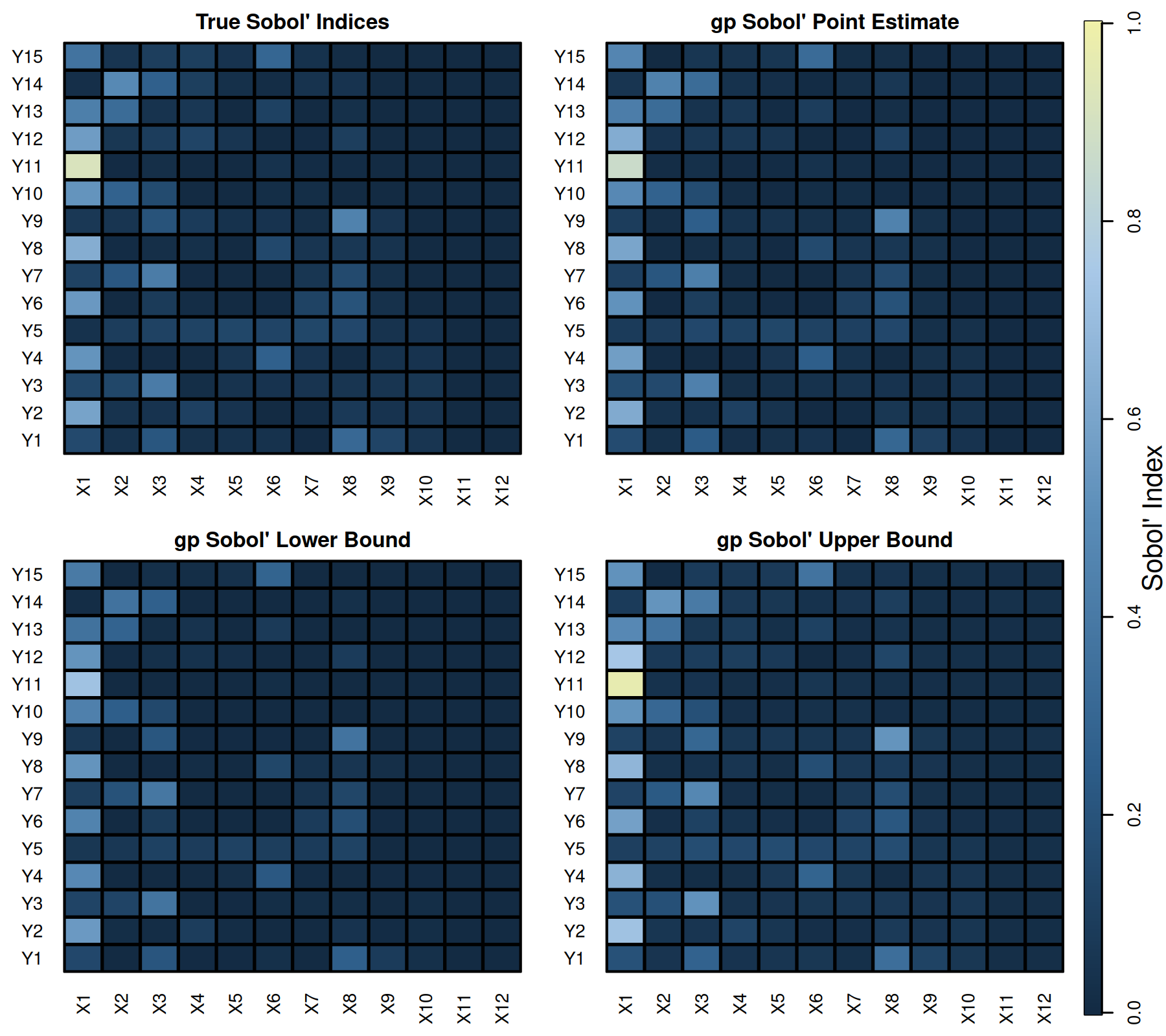}
    \caption{Relative Sobol' indices for the true mean function in the generative process (a), with posterior mean (b) and 95\% posterior credible intervals (c)-(d) for the full GP surrogate.}
    \label{fig:sobol_gp_intervals}
\end{figure}

\begin{figure}[!htbp]
    \centering

    \begin{subfigure}{0.48\textwidth}
        \centering
        \includegraphics[width=\textwidth]{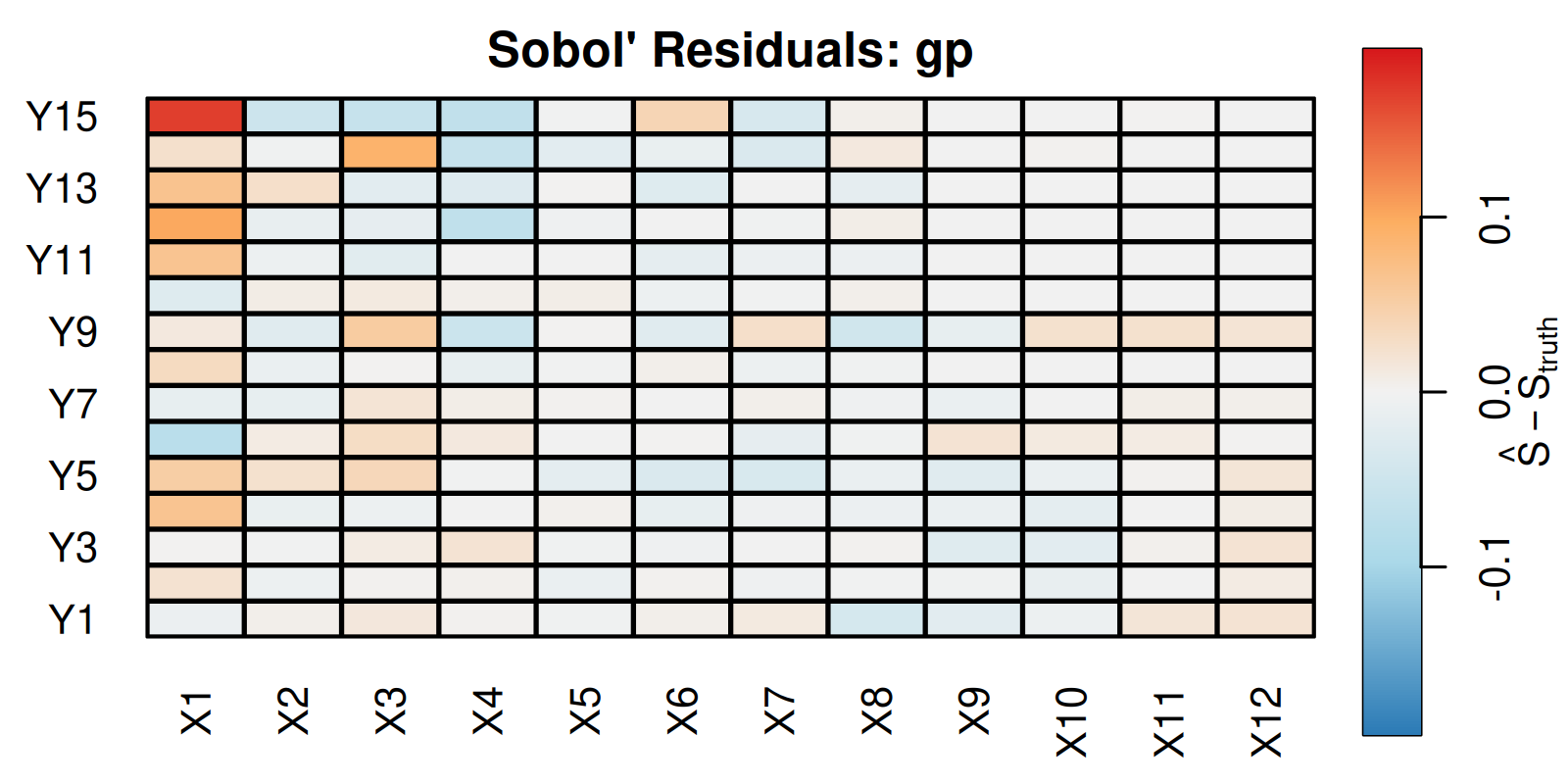}
        \caption{GP.}
        \label{fig:sobol_gp}
    \end{subfigure}
    \hfill
    \begin{subfigure}{0.48\textwidth}
        \centering
        \includegraphics[width=\textwidth]{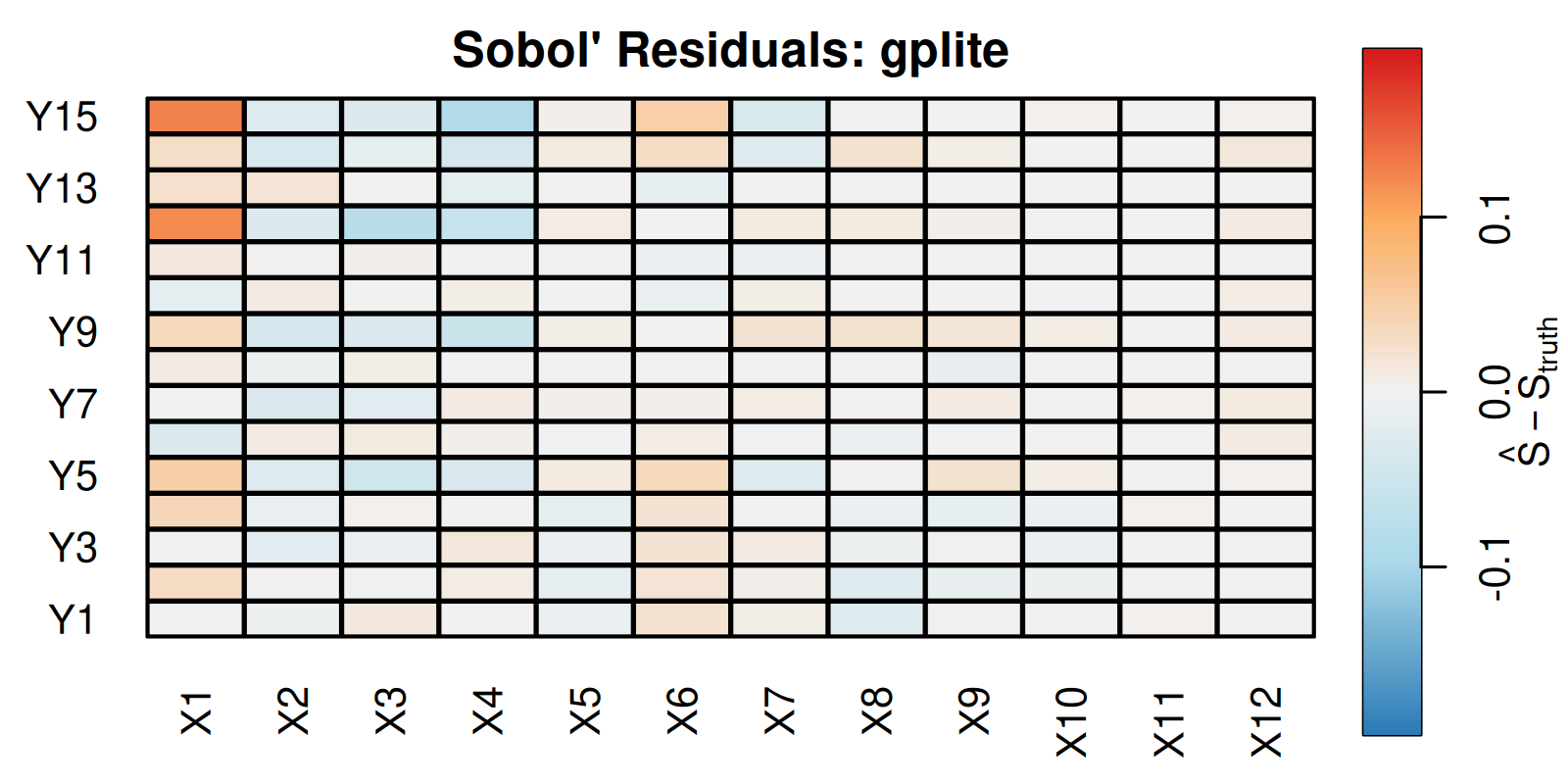}
        \caption{GPLite.}
        \label{fig:sobol_gplite}
    \end{subfigure}

    \vspace{0.75em}

    \begin{subfigure}{0.48\textwidth}
        \centering
        \includegraphics[width=\textwidth]{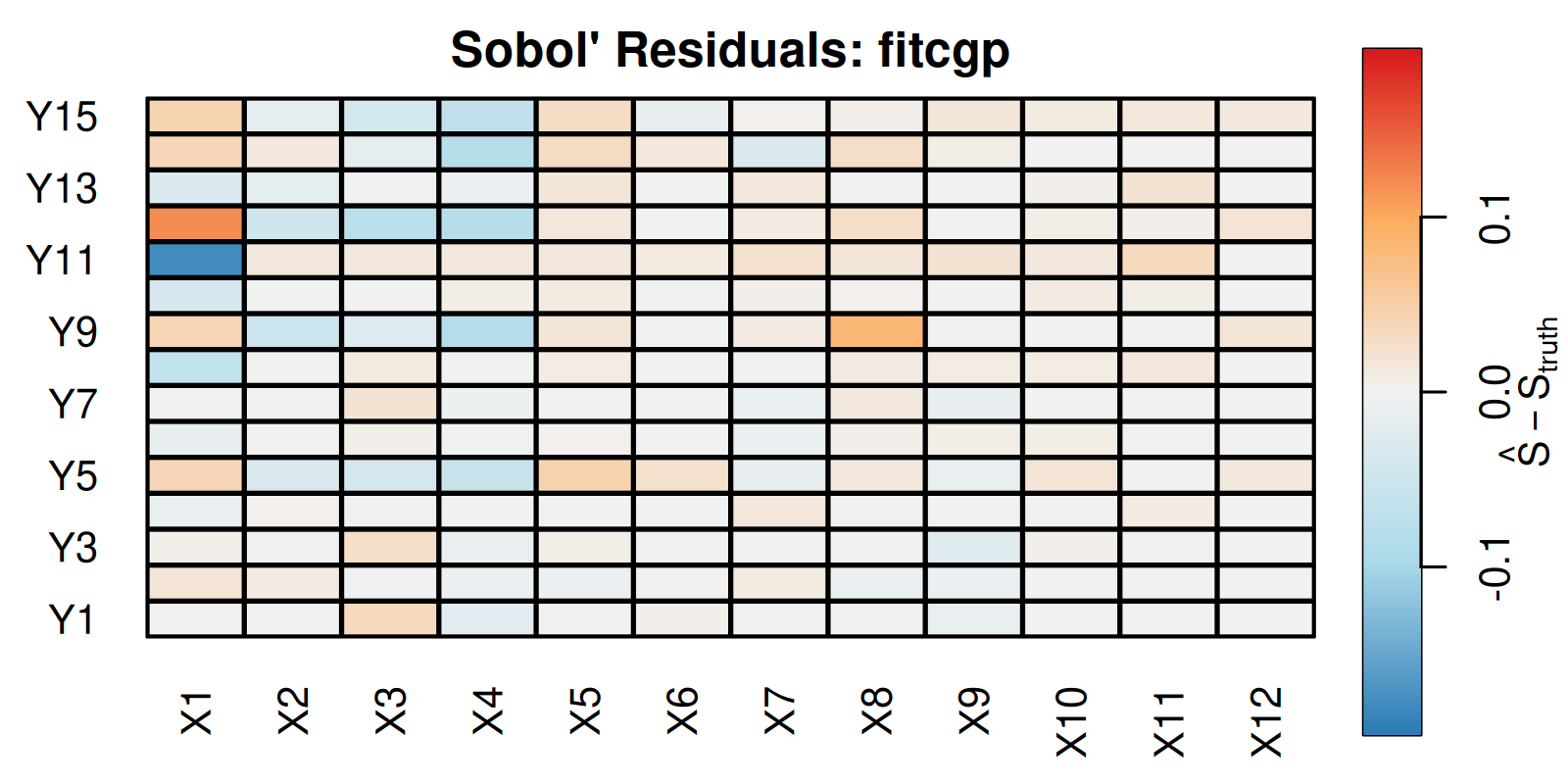}
        \caption{FitCGP.}
        \label{fig:sobol_fitcgp}
    \end{subfigure}
    \hfill
    \begin{subfigure}{0.48\textwidth}
        \centering
        \includegraphics[width=\textwidth]{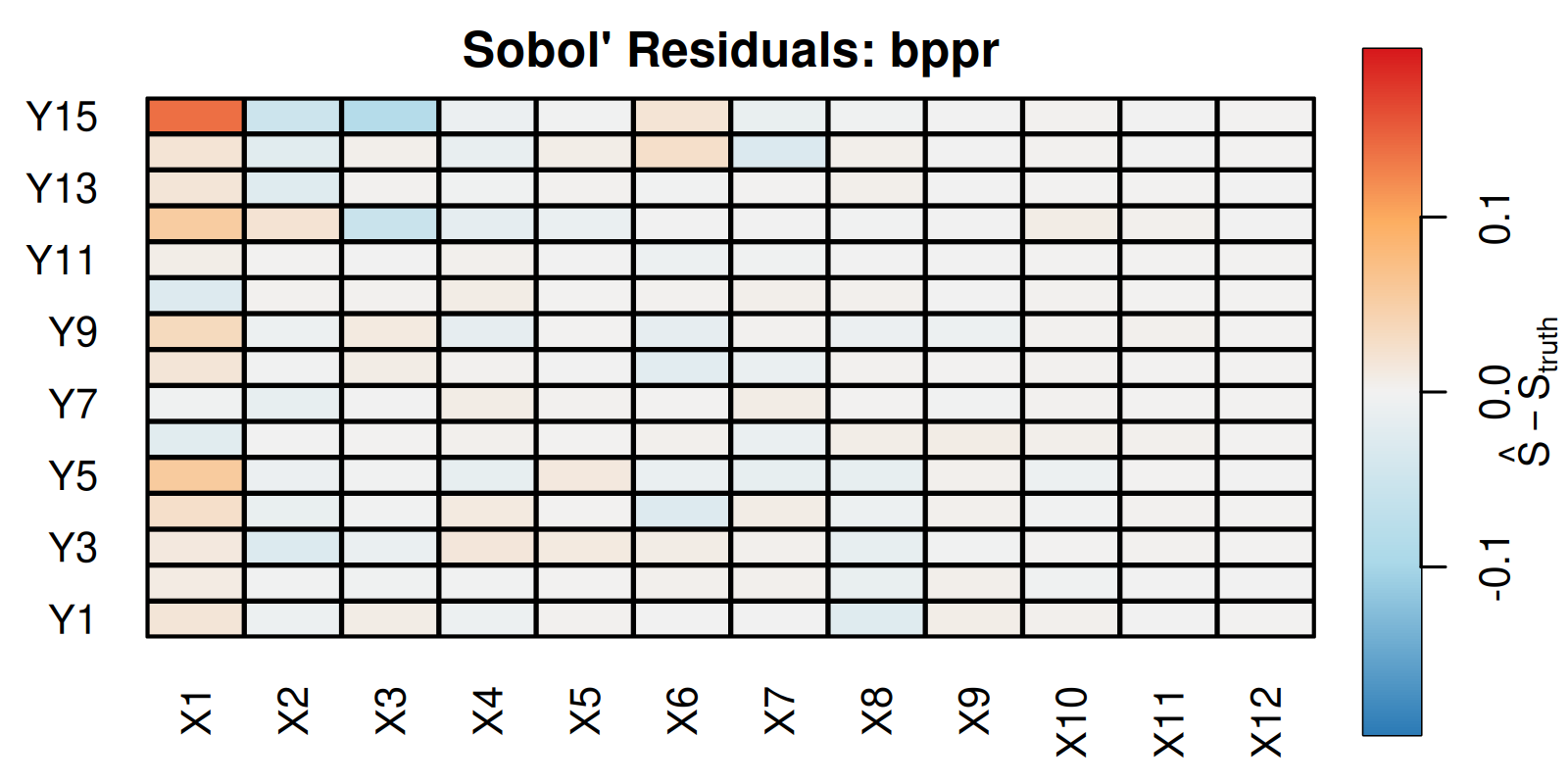}
        \caption{BayesPPR.}
        \label{fig:sobol_bppr}
    \end{subfigure}

    \vspace{0.75em}

    \begin{subfigure}{0.48\textwidth}
        \centering
        \includegraphics[width=\textwidth]{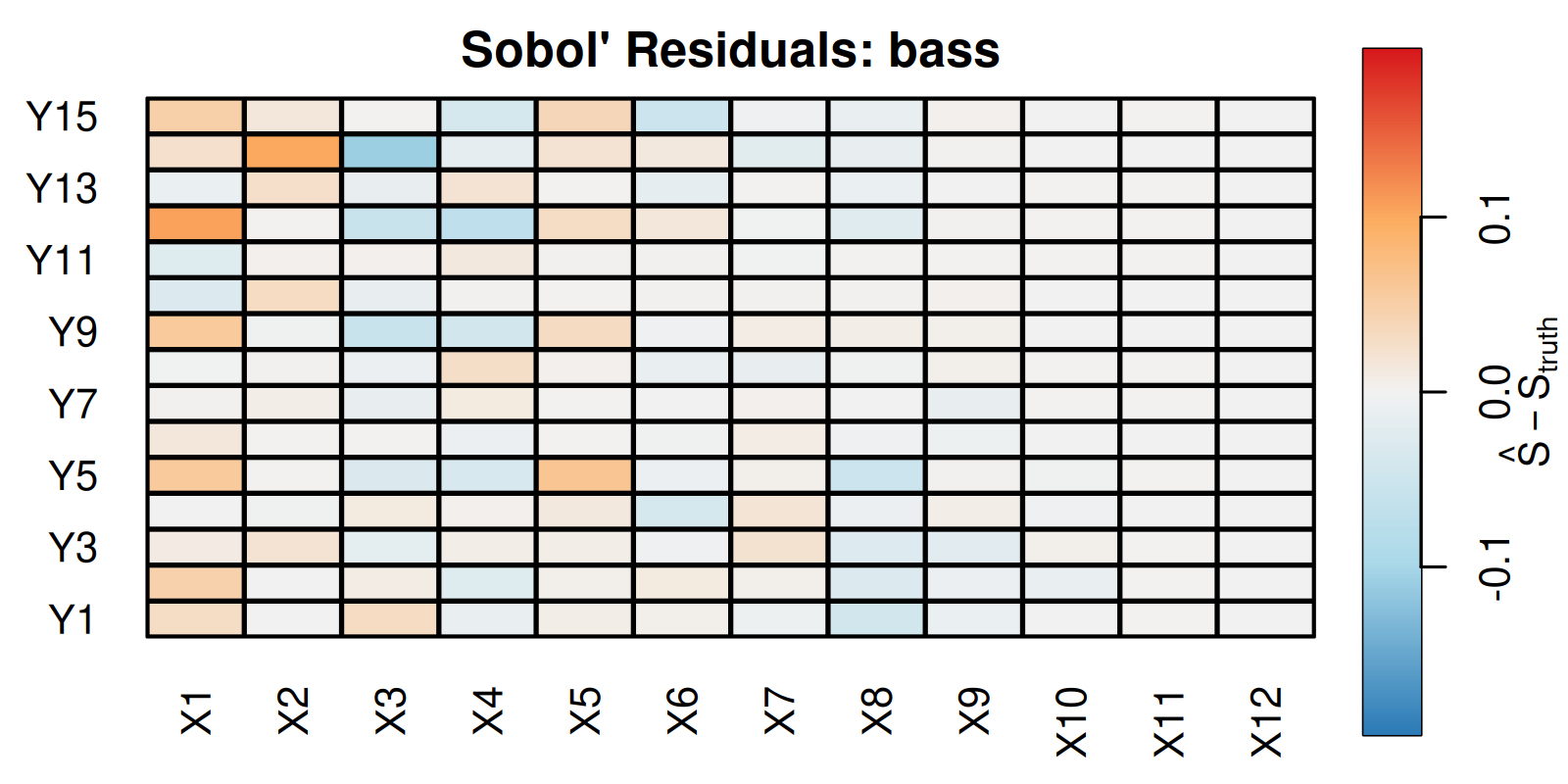}
        \caption{BASS.}
        \label{fig:sobol_bass}
    \end{subfigure}
    \hfill
    \begin{subfigure}{0.48\textwidth}
        \centering
        \includegraphics[width=\textwidth]{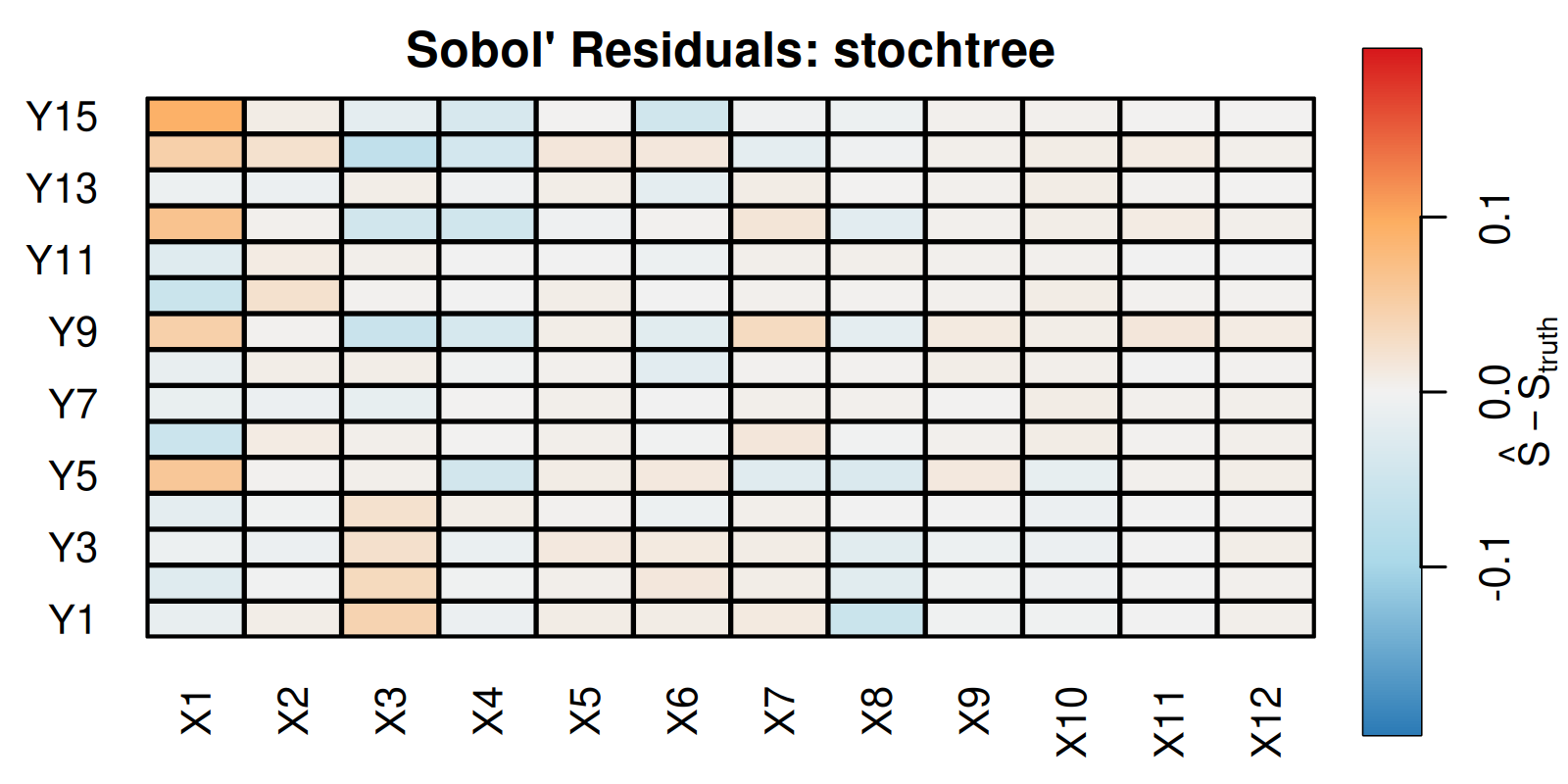}
        \caption{Stochtree.}
        \label{fig:sobol_stochtree}
    \end{subfigure}

    \vspace{0.75em}

    \begin{subfigure}{0.48\textwidth}
        \centering
        \includegraphics[width=\textwidth]{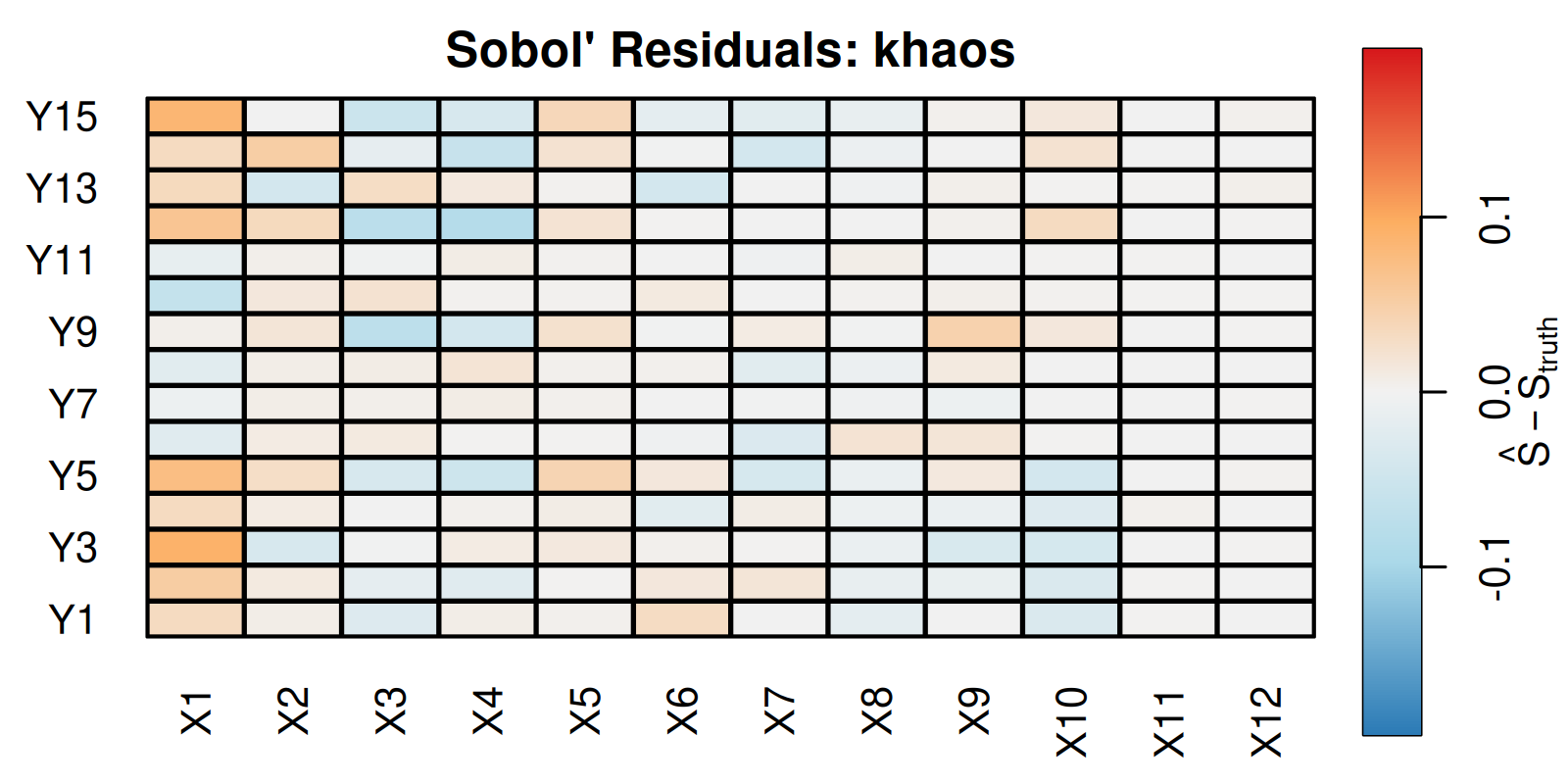}
        \caption{Khaos.}
        \label{fig:sobol_khaos}
    \end{subfigure}
    \hfill
    \begin{subfigure}{0.48\textwidth}
        \centering
        \includegraphics[width=\textwidth]{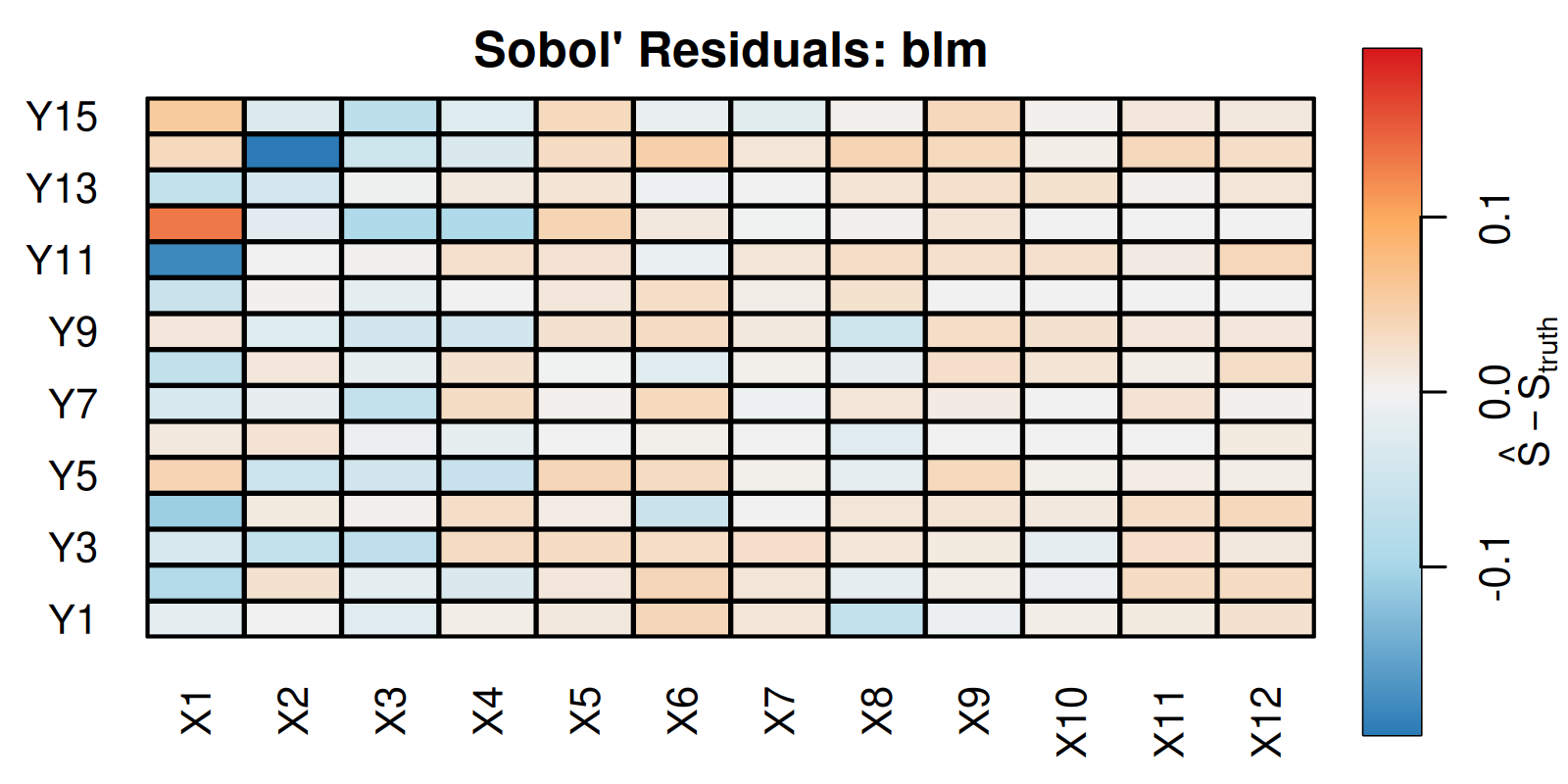}
        \caption{BLM.}
        \label{fig:sobol_blm}
    \end{subfigure}

    \caption{Residuals of the Sobol' sensitivity analysis for the synthetic example. Panels (a)-(h) show the difference from the reference Sobol' indices obtained from the true data-generating mechanism, for each fitted surrogate model.}
    \label{fig:sobol}
\end{figure}

We now apply the fitted surrogates to demonstrate two downstream analyses: Sobol' sensitivity analysis and Bayesian model calibration. For the sensitivity analysis, Figure~\ref{fig:sobol_gp_intervals} displays relative Sobol' indices for the full GP surrogate, with panel (a) showing the gold standard obtained by drawing a Saltelli sample over the input space and propagating it through the true data-generating mechanism, panel (b) showing the posterior mean from the GP, and panels (c) and (d) showing 95\% posterior credible intervals. Figure~\ref{fig:sobol} shows the differences between the gold-standard Sobol' indices and those obtained from each surrogate model by propagating the same Saltelli sample through the fitted surrogate models. Although there are minor differences among methods, all recover the main qualitative structure of the true Sobol' indices and correctly identify the important inputs for the various outputs, and all result in closer approximations than the baseline BLM. 

For calibration, we treat the $n = 500$ responses as simulation outputs and embed each fitted surrogate in a modular Bayesian calibration framework. The goal is to infer a single randomly generated input vector $\bm{\theta}$, which we treat as the true calibration parameter, from a corresponding randomly generated response:

\[
    \bm{y}_{\bm{\theta}} = \Phi f(\bm{\theta}) + \bm{\epsilon},
\]
which we treat as the experimental observation under the assumption of zero model discrepancy. Inference is performed by assigning a uniform prior distribution over the range of the inputs in the training data, and sampling from the posterior distribution with an adaptive Metropolis-Hastings algorithm implemented in the \texttt{impala} package in \textsf{R} \citep{rImpala}, which is designed for Bayesian calibration problems with multivariate responses, as in \cite{francom2025juq}.

\begin{figure}[!htbp]
    \centering
    \includegraphics[width=1.0\textwidth]{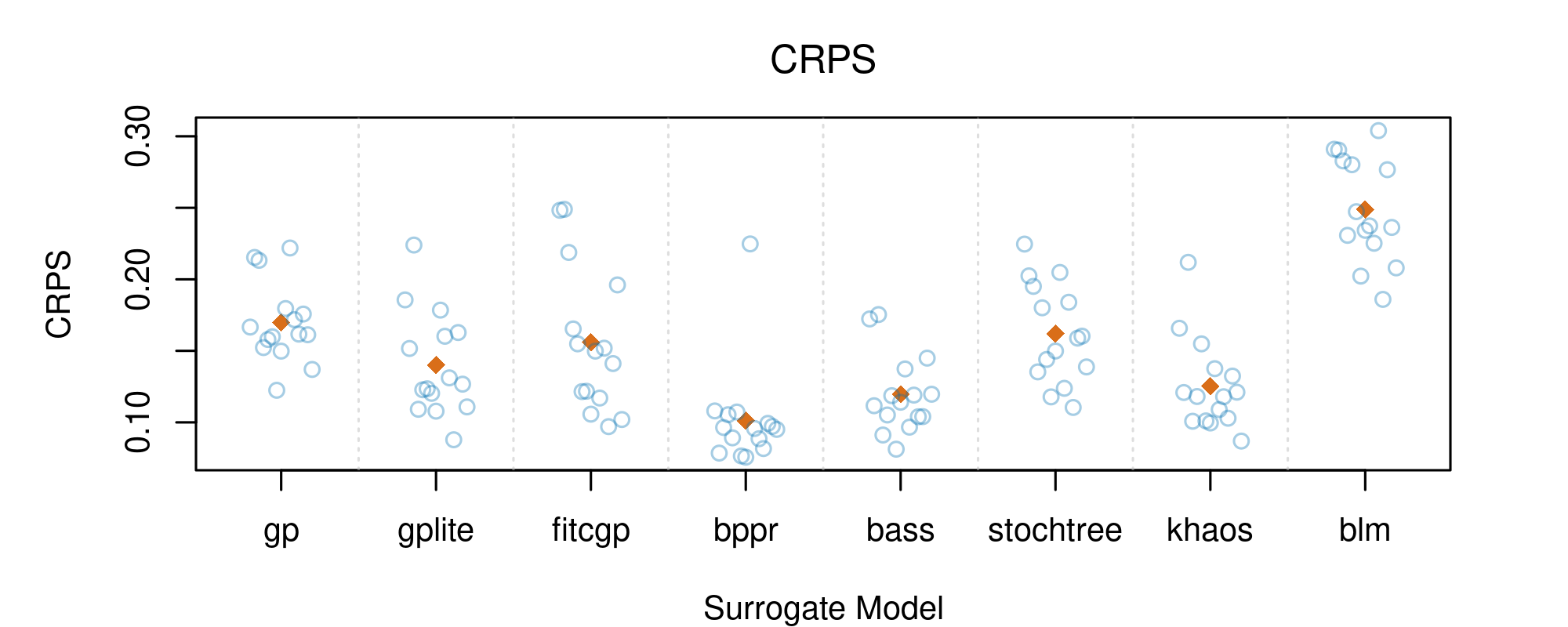}
    \caption{Continuous ranked probability scores (CRPS) for the posterior predictive distributions under Bayesian calibration across surrogate models in the simulated example. Lower values indicate better predictive performance for the calibration response.}
    \label{fig:postpredcrps}
\end{figure}

In Figure~\ref{fig:postpredcrps}, the open circles show the CRPS of the posterior predictive distribution of each component of the true response $\bm y_{\bm\theta}$ used for calibration (in order of response index), with the overall mean CRPS indicated with orange diamonds. BayesPPR achieves the lowest CRPS values, followed by BASS and Khaos, with all surrogates performing substantially better than the BLM baseline. Notably, the results here differ from those in the cross-validation study, where the full GP attained lower predictive CRPS. This is because the value of $\bm\theta$ that was randomly selected for calibration lies in an area of the parameter space where all methods predict with high accuracy. Therefore, methods with narrower intervals around these estimates tend to result in lower CRPS.

\begin{figure}[!htbp]
    \centering
    \includegraphics[width=1.0\textwidth]{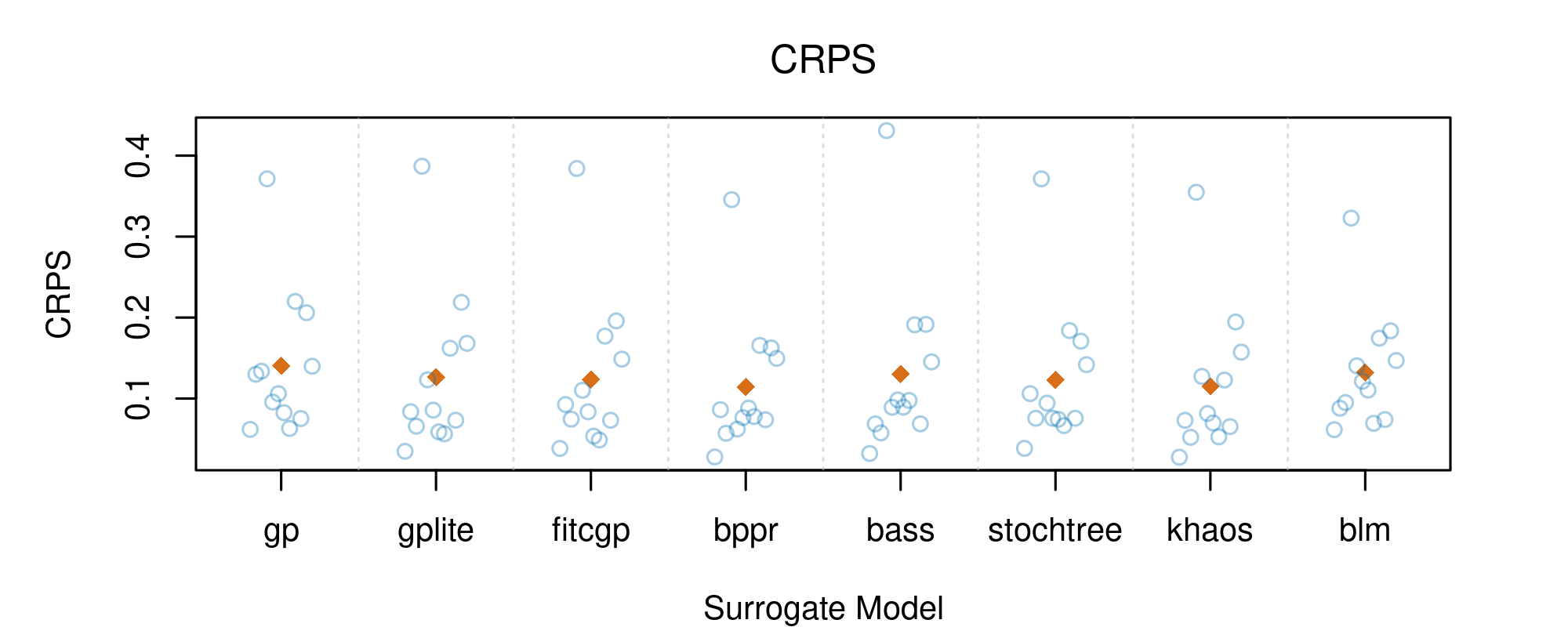}
    \caption{Continuous ranked probability scores (CRPS) for the posterior distributions of the calibration parameters across surrogate models in the simulated example. Lower values indicate better recovery of the true calibration input.}
    \label{fig:postcalcrps}
\end{figure}

Finally, to evaluate the ability of each method to estimate the posterior distribution of the calibration input $\bm \theta$, we compute the CRPS of the posterior distribution for each component of $\bm \theta$ relative to its true value. Figure~\ref{fig:postcalcrps} summarizes these results, ordered from left to right by input variable index, with the mean CRPS across variables again shown in orange diamonds. BayesPPR results in the lowest CRPS, on average, though no surrogate clearly stands out as substantially better or worse than the others for this example.

\subsection{Example Real Data Analysis}\label{sec:application}

We now provide a demonstration of implementing the surrogate models reviewed above to calibrate the equation of state (EoS) of tin using data from pulsed magnetic field experiments \citep{brown2023} conducted using Sandia National Laboratories' Z machine, a pulsed-power driver capable of delivering very large electric currents over short time scales. The goal is to estimate $p=11$ input parameters governing compressibility, that is, the relationship between pressure and density, in order to better understand how materials respond under extreme pressures. 
The tin experiments are designed to investigate both the strength and the phase-transition kinetics of a two-phase system consisting of the ambient beta phase and the high-pressure gamma phase. The parameters $b_{yfac}$ and $g_{yfac}$ relate to the yield strength of the respective phases. In addition, each phase transformation is governed by two kinetics parameters, $(B,\nu)$. Specifically, $B_{bg}$ and $\nu_{bg}$ govern the forward beta-gamma transformation, while $B_{gb}$ and $\nu_{gb}$ govern the reverse transition. Together with the remaining material parameters, these make up the inputs used for surrogate-model construction and calibration. Additional discussion of these material parameters is provided in \cite{met12111844}, and scalar Bayesian calibration for related data is presented in \cite{tin_llnl}.

\begin{figure}[!htbp]
	\centering
	\includegraphics[width=5in]{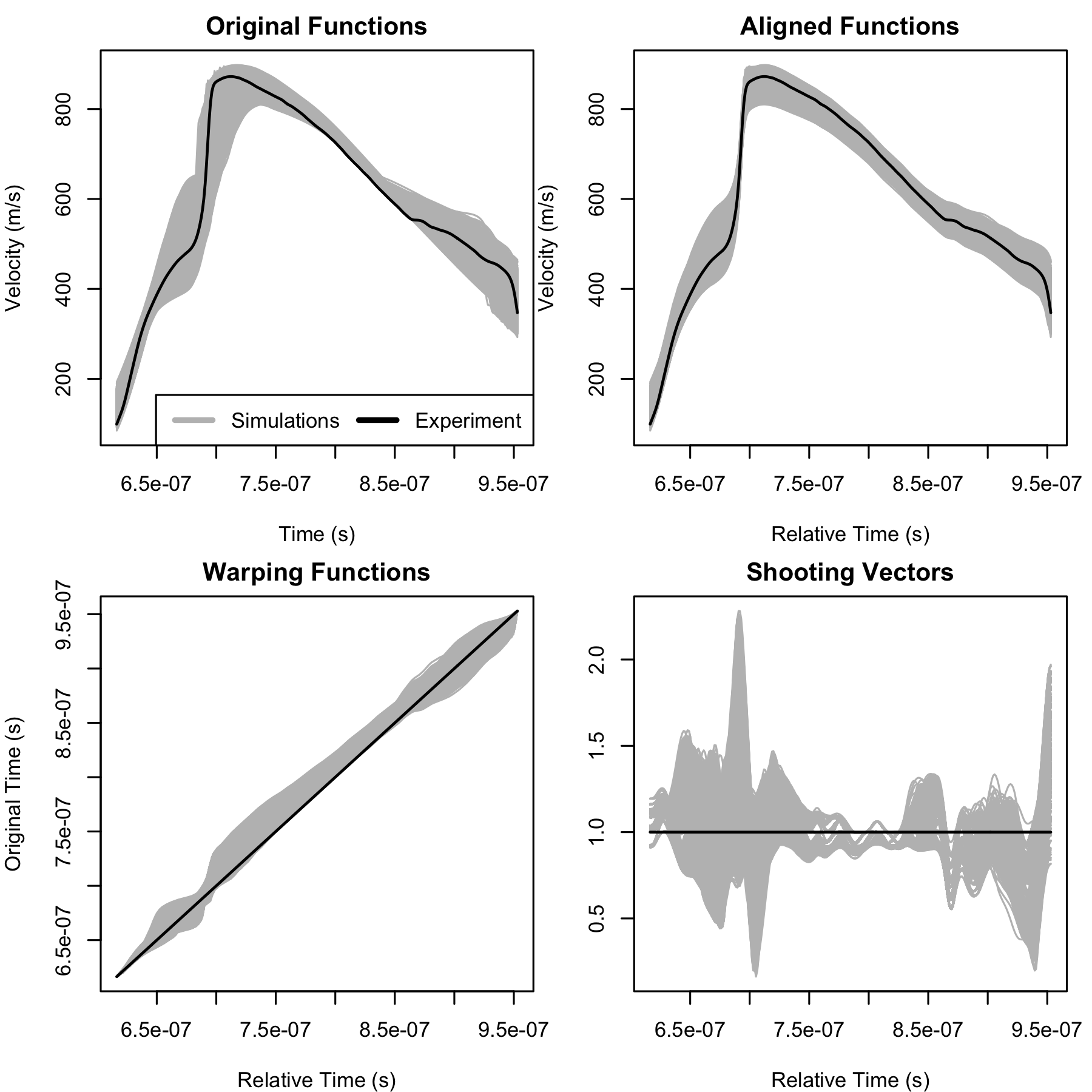}
	\caption{Original simulated and experimental data (top left), with corresponding aligned functions (top right), warping functions (bottom left), and shooting vectors (bottom right) for the Z-machine tin data.}
	\label{fig:tin_data}
\end{figure}

The computer model was evaluated at $n=5499$ distinct input settings, selected using a Latin hypercube sample (LHS) over the 11-dimensional parameter space. The model output consisted of material velocity as a function of time. Corresponding experimental observations were obtained from a Z-machine experiment, and both the simulated and experimental functional outputs were down-sampled to a common grid of $q=300$ time points. Following the approach of \cite{francom2025juq}, and using the \texttt{fdasrvf} package in \textsf{R} \citep{fdasrvf}, the simulated functional outputs were aligned to the experimental data, yielding both aligned functions and warping functions defined relative to the experimental observation. Shooting vectors were then computed from the warping functions, and separate surrogate models were constructed for the aligned computer model outputs and the corresponding shooting vectors, using the basis projection approach outlined in Section~\ref{sec:surr_HD}. Figure~\ref{fig:tin_data} displays the original velocity functions, with the corresponding aligned and warping functions, and shooting vectors, for these data.

As in the simulated example, we conducted a 5-fold cross-validation study using the same set of Bayesian surrogate models, excluding the full GP with MCMC, which does not scale to data of this size. Convergence of the surrogate model MCMC chains is evaluated in Appendix C. Figure~\ref{fig:rmse_tin} reports the RMSE for each emulator, with individual folds again shown as open circles and the mean across folds shown as diamonds. The top and bottom panels correspond to the aligned and warping functions, respectively. BayesPPR achieves the lowest RMSE for both outputs and therefore provides the best predictive accuracy for these data.

\begin{figure}[!htbp]
    \centering
    \includegraphics[width=\textwidth]{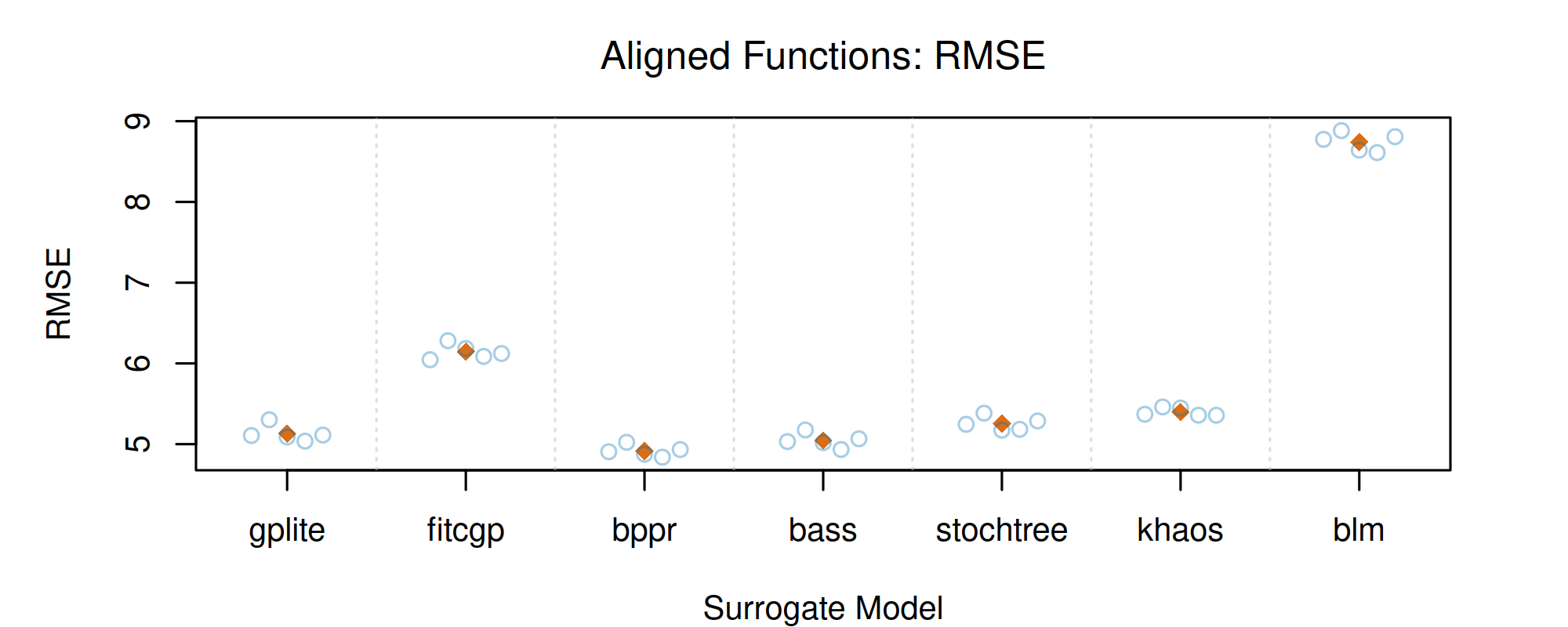} \\
    \includegraphics[width=\textwidth]{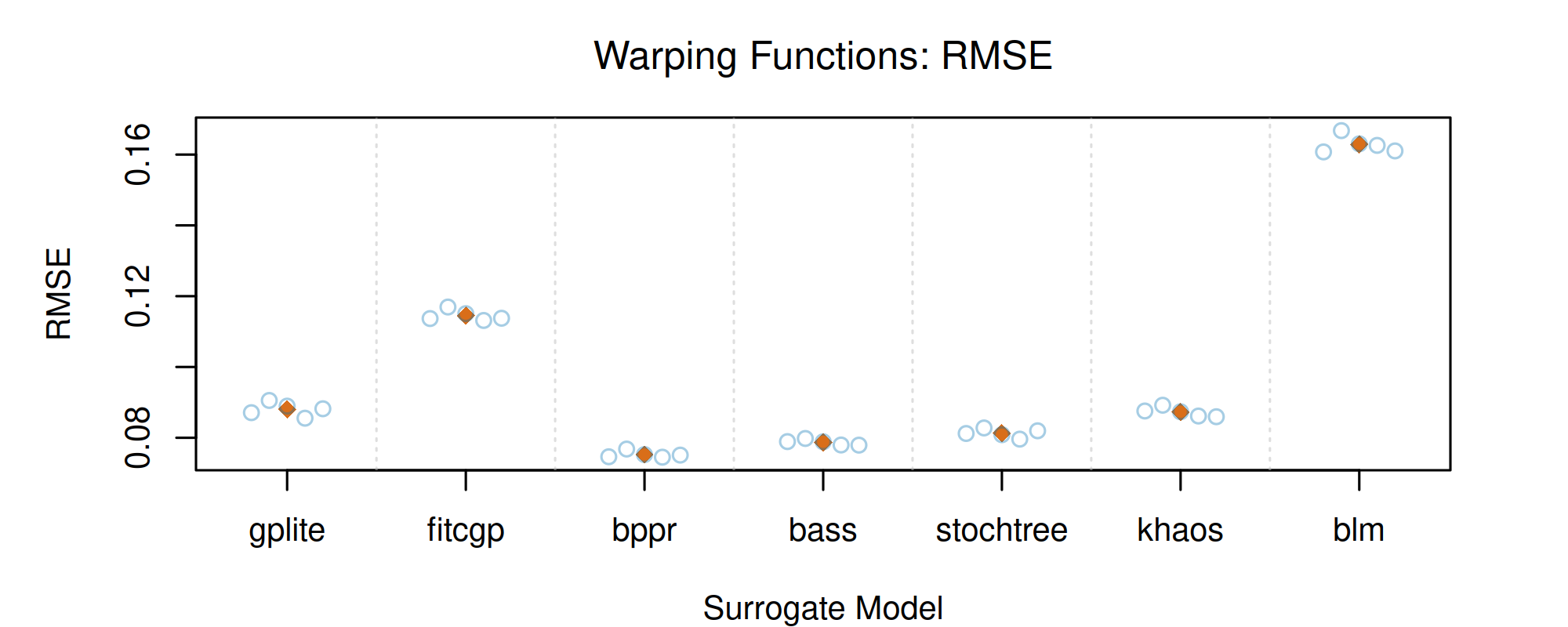}
    \caption{RMSE for aligned functions (top) and warping functions (bottom) for the tin data.} 
    \label{fig:rmse_tin}
\end{figure}

\begin{figure}[!htbp]
    \centering
    \includegraphics[width=\textwidth]{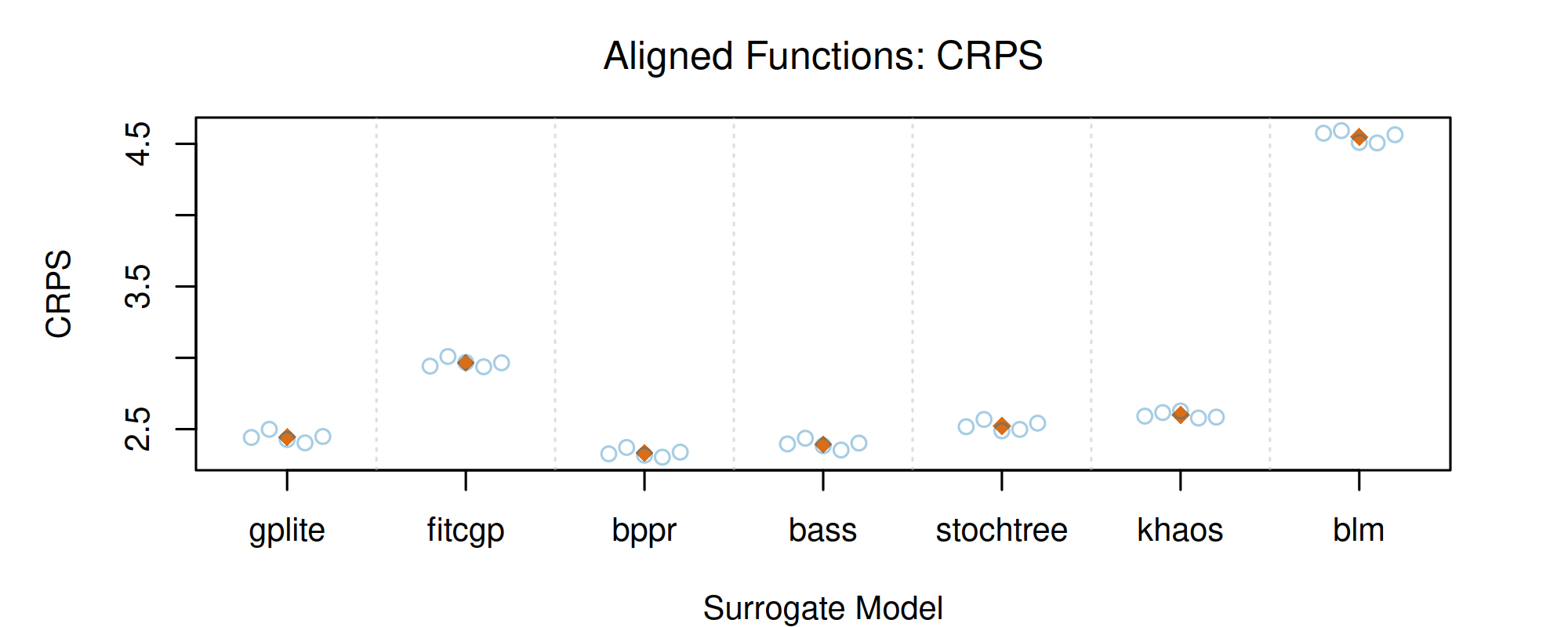} \\
    \includegraphics[width=\textwidth]{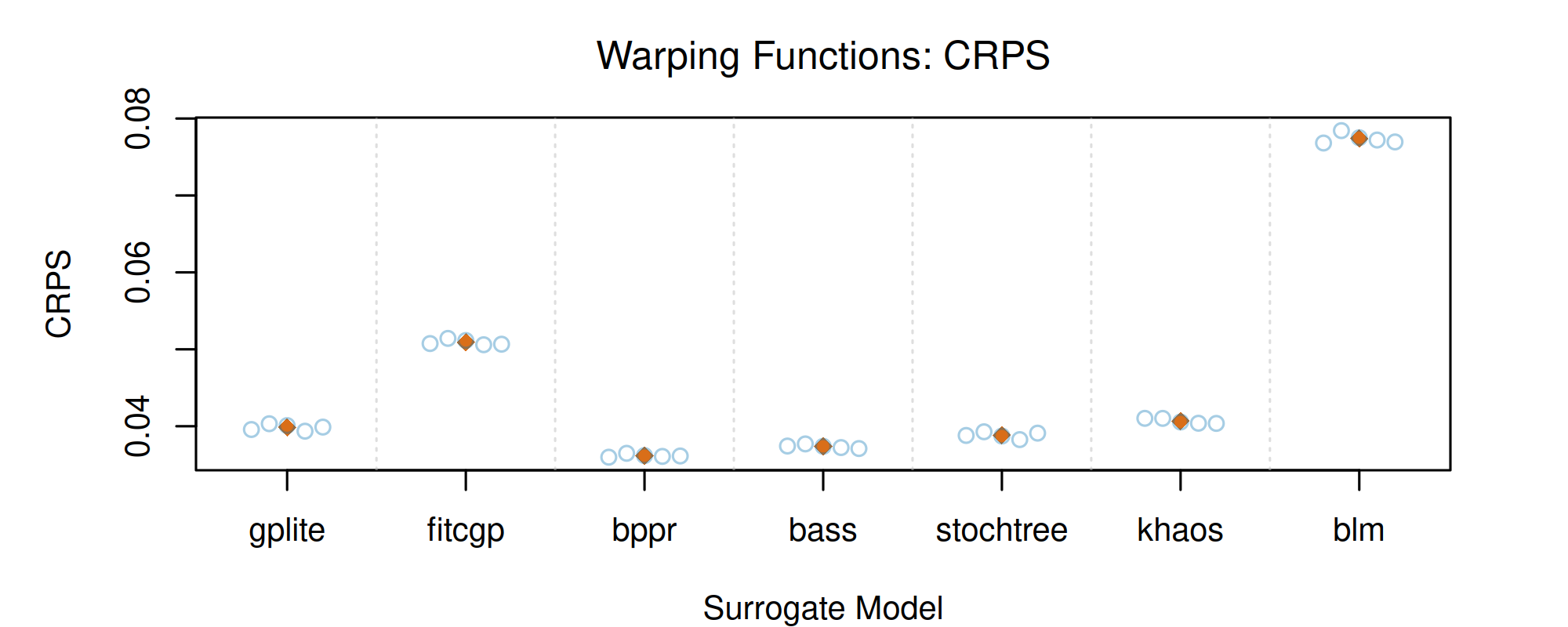}
    \caption{Tin data. CRPS for aligned functions (top) and warping functions (bottom).}
    \label{fig:crps_tin}
\end{figure}

Figure~\ref{fig:crps_tin} reports CRPS across folds to assess UQ, with the top panel again corresponding to the aligned functions and the bottom panel corresponding to the warping functions. BayesPPR attains the lowest CRPS for both the aligned functions and warping functions. The inducing-point method FitCGP appears to be less suitable for these data, likely because the inducing point approximation is too coarse to capture the complex nature of the physics involved, whereas the GPLite variational approximation method performs competitively. Again, we emphasize that these results do not generalize across datasets, and a careful cross-validation study should be performed for each new dataset in practice. See \cite{rumsey2025all} for a demonstration of this principle across a wide range of surrogate models and datasets.

\begin{figure}[!htbp]
    \centering
    \includegraphics[width=1.0\textwidth]{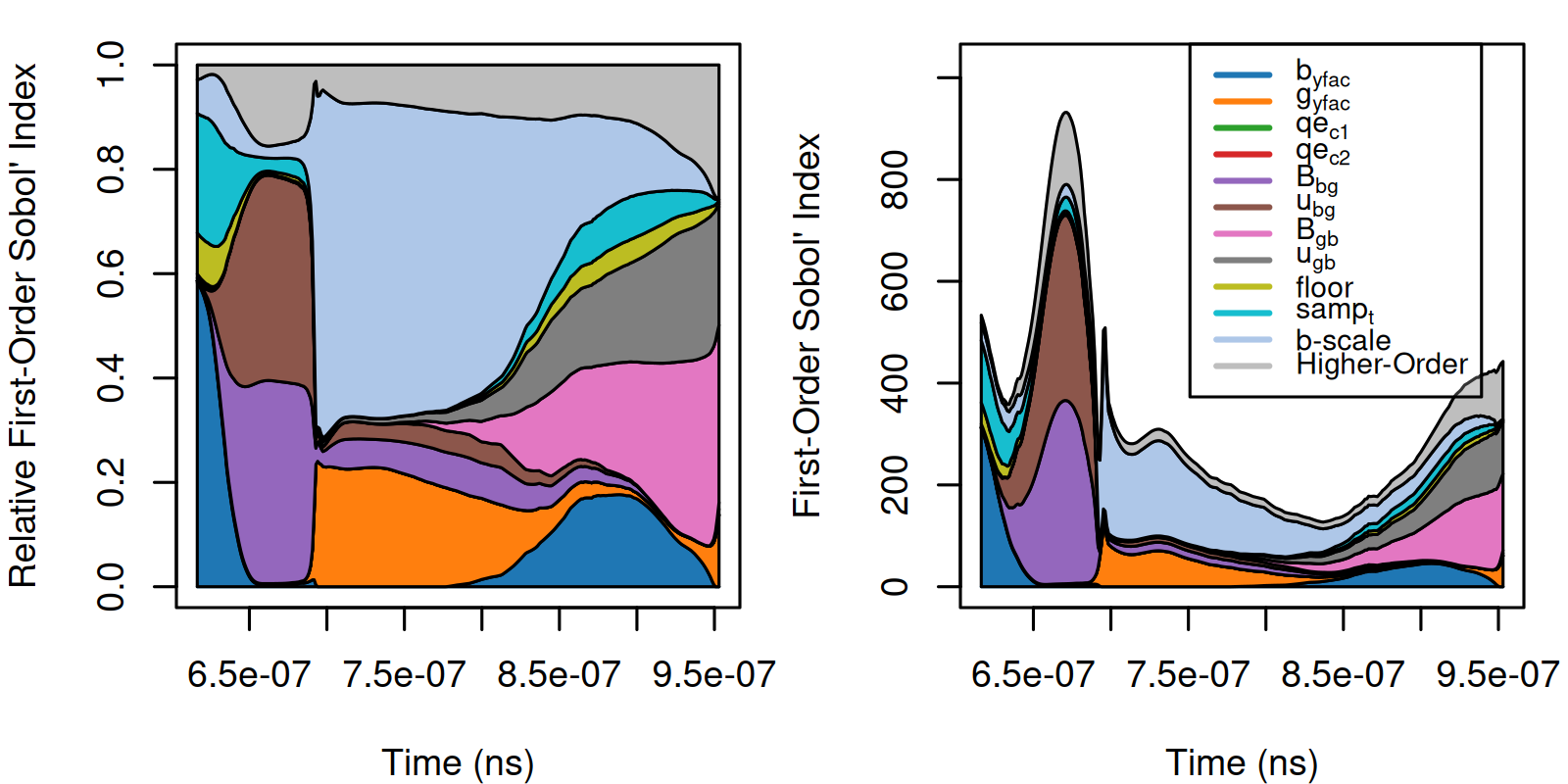} \\
    \includegraphics[width=1.0\textwidth]{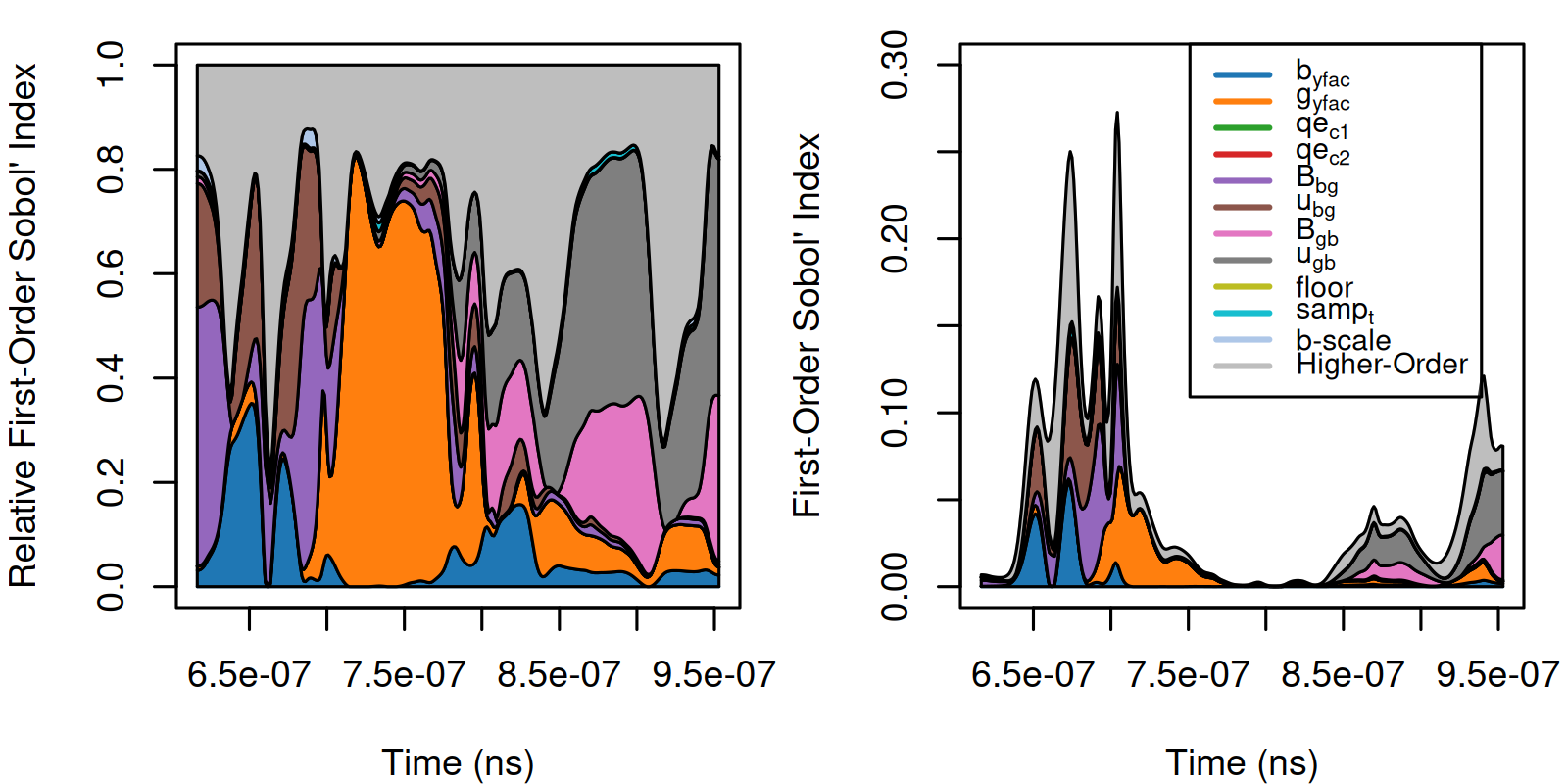}
    \caption{BayesPPR Sobol' sensitivity analysis for aligned functions (top) and warping functions (bottom) for the tin data.}
    \label{fig:Sobol_tin_bppr}
\end{figure}

With these cross-validation results in mind, we restrict the downstream analyses to BayesPPR. Figure~\ref{fig:Sobol_tin_bppr} shows the Sobol' indices for the aligned functions (top) and warping functions (bottom) as functional pie charts. For both sets of surrogates, more than 80\% of the explained variance is attributable to main effects. This representation also shows how the relative importance of the parameters varies over the functional domain. Overall, b-scale is the most influential parameter for the aligned functions, whereas $g_{fac}$ is the most influential for the warping functions, although the relative contributions of the parameters change over time.

Using the BayesPPR surrogate models for both the aligned and warping functions, we performed EFBMC as described in Section~\ref{sec:efbmc}, and in greater detail in \cite{francom2025juq}, assuming no model discrepancy. The calibration was run with 20 parallel chains, including tempering and decorrelation steps. Examination of the chains indicated stable behavior after 2,000 MCMC iterations, and the final 5,000 samples, thinned by a factor of 5, were retained for further analysis.

\begin{figure}[!htbp]
	\centering
	\includegraphics[width=4.5in]{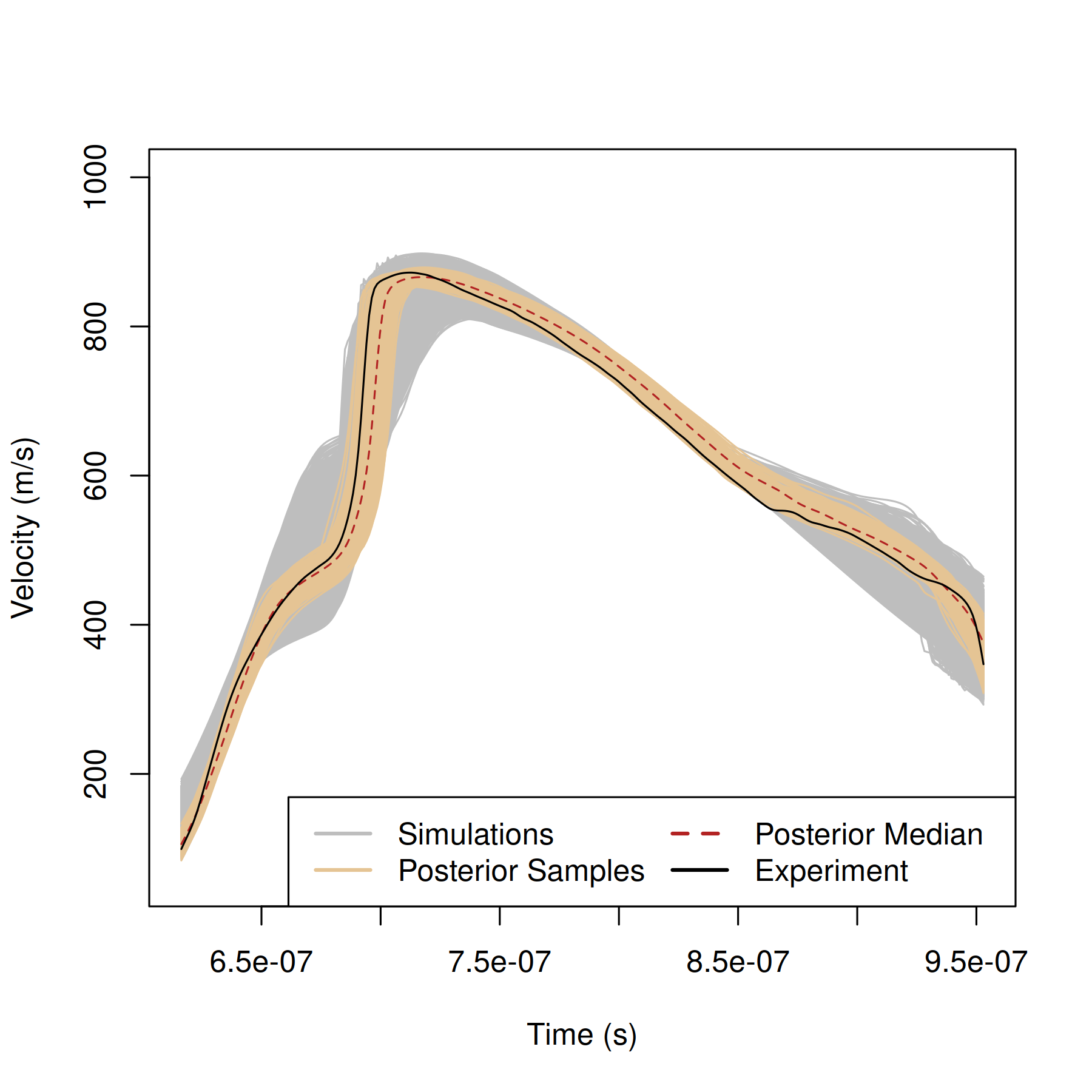}
	\caption{Experimental velocities of tin, compared with the full simulated training set, samples from the posterior predictive distribution from the elastic functional Bayesian model calibration, and the corresponding posterior median.}
	\label{fig:post_cal_yobs_bpprtin}
\end{figure}

Figure~\ref{fig:post_cal_yobs_bpprtin} compares the experimental curve with calibrated surrogate predictions, overlaid on the simulation ensemble, which represents the prior, to show the concentration induced by calibration. To generate these predictions, posterior draws of the input parameters were propagated through the surrogate models for both the aligned functions and shooting vectors. The resulting outputs were then combined to obtain samples in the original function space. The dashed line indicates the posterior median prediction. Overall, the posterior predictive samples follow the experimental curve reasonably well, suggesting agreement between the calibrated model and the observed data.

\begin{figure}[!htbp]
	\centering
\includegraphics[width=1.0\textwidth]{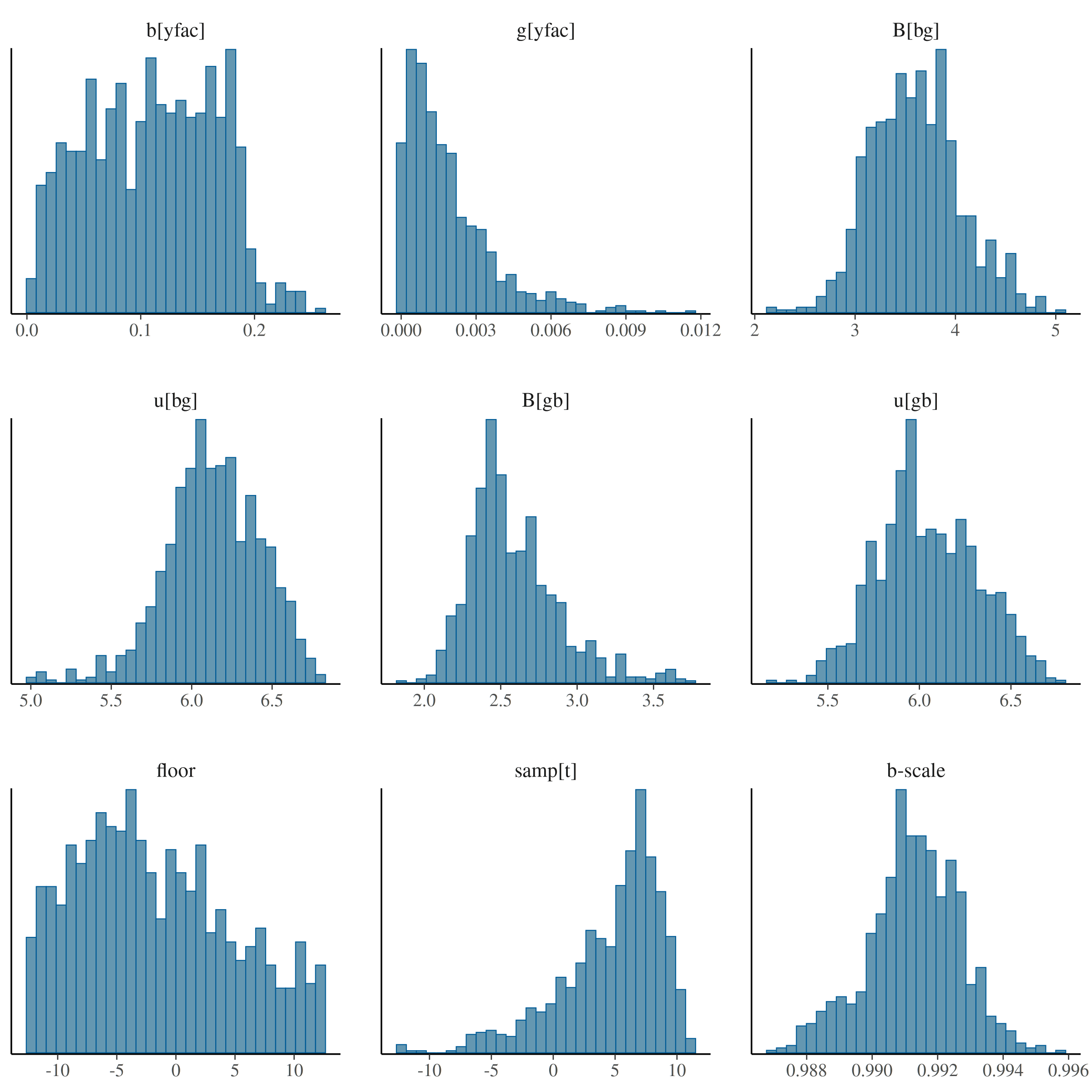}

	\caption{Posterior histograms of calibration parameters for the tin example.}
	\label{fig:post_cal_hist_bpprtin}
\end{figure}

Figure~\ref{fig:post_cal_hist_bpprtin} presents the posterior distributions of the calibration parameters. The posterior for b-scale is relatively concentrated, consistent with its strong influence in the Sobol' analysis, while $\nu_{gb}$ exhibits clear multimodal behavior. Overall, these posteriors appear more informative than those reported in \cite{tin_llnl}, likely because the present calibration uses the full functional response rather than reducing to a scalar summary.

\section{Discussion}\label{sec:summary}

This review provides a high-level overview of widely used Bayesian surrogate modeling methods for high-dimensional computer simulations. We first situated the topic within the broader surrogate modeling literature and discussed the main advantages of Bayesian surrogates, particularly their ability to naturally propagate predictive uncertainty. We then reviewed two broad classes of Bayesian surrogate models: GP methods and Bayesian basis function regression methods, emphasizing the main strengths, limitations, and practical tradeoffs of each. We did not include Bayesian neural networks in this review, primarily because their UQ is not yet as reliable as that of the other methods and because Bayesian inference for them is challenging in surrogate modeling settings. Within the GP class, we also discussed scalable approximations, giving a general overview of approximate GP methodology and describing inducing-point approaches in slightly greater detail. Among Bayesian basis function methods, we highlighted PCE methods, Bayesian MARS, Bayesian PPR, and BART. We note that the set of methods emphasized here necessarily reflects, to some extent, our own experience and the application domains with which we are most familiar.

We next reviewed strategies for surrogate modeling with high-dimensional responses. These included the use of simplified covariance structures, such as Kronecker formulations when appropriate; parallel partial GP constructions; and multivariate frameworks such as the linear model of coregionalization. We also discussed what is perhaps the most commonly used strategy in practice: dimension reduction over the output domain, followed by separate univariate surrogate models for the associated lower-dimensional basis coefficients. This approach often provides an effective balance between flexibility and computational tractability.

In Section~\ref{sec:use_cases}, we then summarized major application areas for Bayesian surrogate models, including forward UQ, global sensitivity analysis such as Sobol' analysis, and Bayesian model calibration. For calibration, we considered settings with scalar outputs, unstructured high-dimensional outputs, and functional outputs, including calibration under the EFDA framework.

To illustrate the practical implications of surrogate model choice, Section~\ref{sec:sim_example} presented a comparative simulation study within the reduced-basis framework. This study examined several surrogate models using cross-validation and evaluated predictive performance through RMSE, interval coverage, and CRPS. We also assessed downstream performance in common surrogate-enabled tasks, specifically Sobol' sensitivity analysis and Bayesian model calibration, emphasizing that surrogate models must achieve both accurate point prediction and reliable uncertainty propagation.

Section~\ref{sec:application} then included a real-data application based on experimental measurements from Sandia National Laboratories' Z machine and associated computer simulations. Using the EFDA framework, we compared multiple surrogate modeling approaches, identified the best-performing method for these data, and then used that surrogate to conduct Sobol' sensitivity analysis and Bayesian model calibration. To support reproducibility, example scripts are available at \url{https://github.com/sandialabs/EFBMC} in \textsf{R}, MATLAB, and Python. These scripts illustrate Bayesian surrogate modeling and EFBMC on both synthetic examples and a Z-machine calibration problem similar to what we presented here.

We now provide practical guidance on surrogate model selection, high-dimensional output handling, and UQ evaluation, and outline key areas for future research.

For problems with moderate input dimensionality where UQ is a primary objective, GP surrogates remain a reliable choice because of their flexibility and principled predictive uncertainty. However, full Bayesian inference for exact GPs doesn't scale well for moderate-to-large numbers of training simulations. Approximate GP methods offer scalability, though their effectiveness depends on how well the approximation matches the data. Bayesian basis function regression methods, including BART and Bayesian MARS, provide flexible alternatives that can capture complex nonlinearities and interactions and scale well computationally. Currently, Bayesian neural networks are less recommended for surrogate modeling because their UQ is not yet as reliable and inference remains challenging in these settings.

When dealing with high-dimensional outputs, dimension reduction followed by independent univariate surrogate modeling of basis coefficients is a practical and widely used strategy. However, users should be mindful of the limitations of the common assumption of independent residuals, which can oversimplify dependencies in the output space. Where computational resources permit, multivariate or structured covariance models should be explored to better capture these dependencies.

We also emphasize evaluating surrogate models not only on point prediction accuracy but also on UQ metrics such as CRPS. This is especially important when surrogates are used for downstream tasks such as global sensitivity analysis and Bayesian model calibration, where reliable uncertainty propagation is essential.

Substantial progress has been made in surrogate modeling for scalar, functional, and spatial outputs, but many open problems remain. One important direction for future work is the development of frameworks for increasingly complex output types, including tensors, graphs, and other highly structured data. More broadly, there is a need for alternative frameworks for general high-dimensional responses that move beyond the common practice of fitting independent surrogate models to basis coefficients. For example, the widespread assumption of i.i.d.\ Gaussian residuals can oversimplify dependence in the resulting high-dimensional covariance structure. There is also a need for guidance on method selection, including systematic benchmarking efforts and standardized evaluation protocols. While it is clear that different surrogate models perform differently across applications, there has been comparatively little work on identifying which classes of models are best suited to which types of simulators, data regimes, and inferential goals. Continued advances are also needed in computational scalability, predictive accuracy, and fidelity of UQ. In the examples considered here, for instance, no method uniformly achieved ideal predictive performance and perfectly calibrated uncertainty, underscoring the difficulty of high-dimensional surrogate modeling and the opportunity for further methodological development.

In summary, Bayesian surrogate modeling is a central tool for the analysis of computationally expensive simulators, particularly when UQ is a primary objective. As simulation models become increasingly complex and produce richer outputs, developing Bayesian surrogates that are scalable, accurate, and equipped with well-calibrated UQ will remain an important research area with broad impact across the physical sciences and engineering disciplines.

\section*{Disclaimer}
This paper describes objective technical results and analysis. Any subjective views or opinions that might be expressed in the paper do not necessarily represent the views of the U.S. Department of Energy or the United States Government.

This article has been authored by an employee of National Technology \& Engineering Solutions of Sandia, LLC under Contract No. DE-NA0003525 with the U.S. Department of Energy (DOE). The employee owns all right, title and interest in and to the article and is solely responsible for its contents. The United States Government retains and the publisher, by accepting the article for publication, acknowledges that the United States Government retains a non-exclusive, paid-up, irrevocable, world-wide license to publish or reproduce the published form of this article or allow others to do so, for United States Government purposes. The DOE will provide public access to these results of federally sponsored research in accordance with the DOE Public Access Plan https://www.energy.gov/downloads/doe-public-access-plan

\bibliographystyle{plainnat}
\bibliography{bibl}
\end{CJK}
\end{document}